\documentclass[10pt,journal,compsoc]{IEEEtran}
\ifCLASSOPTIONcompsoc
  \usepackage[nocompress]{cite}
\else
  \usepackage{cite}
\fi
\ifCLASSINFOpdf
  \usepackage[pdftex]{graphicx}
  \DeclareGraphicsExtensions{.pdf,.jpeg,.png,.jpg}
\else
  \usepackage[dvips]{graphicx}
  \DeclareGraphicsExtensions{.eps}
\fi
\graphicspath{{figures/}{pictures/}{images/}{./}}

\usepackage{microtype}                 
\PassOptionsToPackage{warn}{textcomp}  
\usepackage{textcomp}                  
\usepackage{mathptmx}                  
\usepackage{times}                     
\usepackage{cite}                      
\usepackage{tabu}                      
\usepackage{booktabs}                  
\usepackage{multirow}
\usepackage{chngpage}
\usepackage{hyperref}
\usepackage{threeparttable}
\usepackage{color, soul}
\usepackage{diagbox}
\usepackage{dblfloatfix}    
\usepackage{float}
\usepackage{setspace}
\usepackage{enumitem}
\usepackage{cleveref}
\usepackage[bottom]{footmisc}

\setitemize{noitemsep,topsep=1pt,parsep=0pt,partopsep=0pt}
\definecolor{myblue}{RGB}{52, 152, 219}
\definecolor{myorange}{RGB}{230, 126, 34}
\definecolor{mygreen}{RGB}{46, 204, 113}
\definecolor{myred}{RGB}{255, 0, 0}

\begin{document}
\begin{spacing}{0.98}
%
\title{MARVisT: Authoring Glyph-based Visualization in Mobile Augmented Reality}
%
%
%
%

\author{Chen Zhu-Tian, 
	    Yijia Su,
	    Yifang Wang,
	    Qianwen Wang,
	    Huamin Qu,
	    Yingcai Wu
	    
\IEEEcompsocitemizethanks{
  \IEEEcompsocthanksitem C. Zhu-Tian is with the State Key Lab of CAD\&CG, Zhejiang University and Hong Kong University of Science and Technology. A part of this work was done when Chen Zhu-Tian  was a visiting student supervised by Yingcai Wu at the State Key Lab of CAD\&CG, Zhejiang University.
    	E-mail: zhutian.chen@connect.ust.hk.
  \IEEEcompsocthanksitem Y. Su, Y. Wang, Q. Wang and H. Qu are with Hong Kong University of Science and Technology. E-mail: \{yijiasu, yifangwang, qwangbb\}@connect.ust.hk, huamin@cse.ust.hk.
  \IEEEcompsocthanksitem Y. Wu is with the State Key Lab of CAD\&CG, Zhejiang University. Yingcai Wu is the corresponding author. E-mail: ycwu@zju.edu.cn.
 }
	    
}

%
%

\markboth{Journal of \LaTeX\ Class Files,~Vol.~14, No.~8, August~2015}%
{Shell \MakeLowercase{\textit{et al.}}: Bare Demo of IEEEtran.cls for Computer Society Journals}
%



\IEEEtitleabstractindextext{%
\begin{abstract}
  Recent advances in mobile augmented reality (AR) techniques have shed new light on personal visualization 
  for their advantages of fitting visualization within personal routines, 
  situating visualization in a real-world context, 
  and arousing users' interests. 
  However, enabling non-experts to create data visualization 
  in mobile AR environments is challenging 
  given the lack of tools that allow in-situ design 
  while supporting the binding of data to AR content. 
  Most existing AR authoring tools require working on personal computers 
  or manually creating each virtual object and modifying its visual attributes. 
  We systematically study this issue 
  by identifying the specificity of AR glyph-based visualization authoring tool
  and distill four design considerations. 
  Following these design considerations, 
  we design and implement MARVisT, 
  a mobile authoring tool 
  that  
  leverages information from reality to assist non-experts in
  addressing relationships between data and virtual glyphs, 
  real objects and virtual glyphs, 
  and real objects and data.
  With MARVisT, users without visualization expertise 
  can bind data to real-world objects to create expressive AR glyph-based visualizations rapidly and effortlessly, 
  reshaping the representation of the real world with data. 
  We use several examples to demonstrate the expressiveness of MARVisT.
  A user study with non-experts is also conducted to evaluate the authoring experience of MARVisT.
\end{abstract}

}

\maketitle

\IEEEdisplaynontitleabstractindextext

%
\IEEEpeerreviewmaketitle

\ifCLASSOPTIONcompsoc
\IEEEraisesectionheading{\section{Introduction}\label{sec:introduction}}
\else
\section{Introduction}
\label{sec:introduction}
\fi


\IEEEPARstart{W}{ith} the growing availability of data in our daily life, 
personal visualization (PV),
which aims to help the public better visually explore  
data of interest in a personal context, 
has experienced a recent research surge, 
attracting attention from researchers 
in the visualization and human-computer interaction (HCI) communities~\cite{PV&PVA}.
Unlike traditional visualization applications that are oriented to domain experts, 
PV is less task focused~\cite{PV&PVA, casualInfoVis} and is highly correlated to daily life. 
The growth in mobile augmented reality (AR) techniques (ARKit~\cite{ARKit} on iOS and ARCore~\cite{ARCore} on Android) 
have shed new light on PV.
We argue that mobile AR techniques have great potential 
to facilitate PV.
First, mobile devices, 
which have become indispensable in our daily routines,
represent one of the most available all-in-one platforms to collect, process, and display data related to personal context.
Second, by using the AR techniques, 
users can situate visualization into a real-world setting,
thus enhancing the association between the data being visualized and the real background~\cite{ia-position, sa}.
However, mainstream desktop-based AR content authoring tools (\emph{e.g.}, Unity and Vuforia) 
are not suitable for non-experts to create 
mobile AR visualization. 
Two gaps primarily contribute to such unsuitability:
1) the steep learning curve of these fully-featured tools, 
and 2) the inconvenient offline workflow wherein users must work back-and-forth between mobile devices and desktop PCs.
These characteristics have motivated us to design and develop a mobile authoring tool for the efficient creation of AR visualization.

This research takes the first step towards
enabling non-experts to use mobile devices to create data visualizations in AR environments.
We specifically focus on the authoring of glyph-based data visualization for two reasons.
First, as a type of infographic, 
the advantage of glyph-based visualization
meets the requirements of PV,
namely,
conveying data messages quickly and attractively to the public~\cite{glyph-survey}, 
who usually only have limited experience with data visualization.
Second, glyph-based visualization objectifies each data point as a physical object 
that is suitable to display in three-dimensional (3D) AR environments, 
whereas 3D data charts remain controversial~\cite{tamara-visualization,3dgood}.
We abbreviate the term \emph{AR glyph-based visualization} to \emph{ARGVis} for brevity.

Although various existing works presented methods~\cite{DDG, ivisdesigner, ivolver, lyra} 
to assist users in creating visualizations,
explorations on mobile authoring tools for ARGVis are still scarce.
Numerous open questions remain about creating ARGVis in mobile devices.
To identify the specificity and challenges of authoring ARGVis,
we first compare the abstract workflow of 
traditional visualization authoring tools
with ARGVis authoring tools
to assess the 
similarities and 
differences between them.
Findings reveal that the characteristics of ARGVis are critical in the design of the tools.
Thus,
we situate ARGVis within the context of the literature
to characterize its essential components and the relationships among these components.
Based on the characterization,
we identify that
while traditional 
authoring 
tools focus on relationships 
between data and visual channels,
ARGVis authoring tools should further address the relationships
between reality and virtuality.
Specifically,
an ARGVis authoring tool should
help users 
create visualizations that reveal 
the relationships between
1) data and virtual glyphs,
2) real objects and virtual glyphs,
and 3) real objects and data.
How to support these kinds of authoring remains an open question.
To address this issue, 
we distill and articulate 
a set of four design considerations
based on our analysis and existing authoring tools,
and refine them through several design iterations.
Following these design considerations,
we present MARVisT, 
a \textbf{M}obile \textbf{AR} \textbf{Vis}ualization \textbf{T}ool,
tailored for non-experts to create creative visualizations in AR environments.
Compared with the state-of-art visualization authoring tools, 
MARVisT leverages information from reality to assist users in 
authoring visual mapping, visual scales, and layouts
for creating ARGVis.
To demonstrate the expressiveness and usefulness of MARVisT,
we present examples 
that are difficult and tedious for non-experts to construct in traditional PCs.
We also conduct a user study 
with participants without visualization and AR expertise
to evaluate the authoring experience of MARVisT.

To summarize, our contributions are as follows:
\begin{itemize}
\item A systematical study of the specificities of ARGVis authoring tools and the characterization of ARGVis. 
\item A mobile authoring tool which features context-aware nudging, visual scales synchronization,
and virtual glyphs auto-layout to enable non-experts to create ARGVis.
\item An evaluation with examples and a user study to demonstrate the expressiveness, usability and usefulness of the tool.
\end{itemize}

\vspace{-4mm}
\section{Related Works}
\label{sec:relatedworks}

\subsection{Visualization in Augmented Reality}
Recent years have seen a surge in the popularity of consumer-grade AR devices (\emph{e.g.}, head-mounted displays (HMD), mobile phones, and tablets) 
which has led to an increasing number of studies exploring the visualization and visual analytics based on AR techniques.
In the visualization community, the works related to AR 
can be roughly classified according to the computing modalities that create the AR content, 
namely, HMD systems~\cite{ar-urban-exploded, ar-layout-planning} 
and mobile handheld systems~\cite{sa, sa-ego-centric}.
Currently, HMD-based devices have various deficiencies, 
including narrow field-of-view, low resolution, and inaccurate tracking~\cite{ar-cost}.
Moreover, the HMD devices are still relatively new to most consumers.
Bach et al.~\cite{ar-in-my-hand} reported that participants needed additional time to become familiar with the HMD devices;
otherwise, they would perform worse in the interactive exploration tasks.

As another type of AR device, mobile devices are more widely available and familiar to the masses.
Several works based on mobile AR devices have been introduced for various application scenarios.
MelissAR~\cite{ia-mobile-bee}, a tablet AR system, showing on-site drift information 
to assist beekeepers in understanding bee behaviors.
Based on real-time face recognition, 
Zhao et al.~\cite{sa-ego-centric} developed an iPhone-based ego-centric network AR visualization system 
supporting on-site analysis of temporal co-authoring patterns of researchers.
Nicola et al.~\cite{sa-geology} proposed a sketch-like AR visualization to support the field workflow of geologists. 
Despite the extensive use of mobile AR to present data in various professional fields,
studies have rarely applied such techniques in casual scenarios,
such as people's daily lives (\emph{e.g.}, using mobile AR for shopping assistance~\cite{sa}), 
which is one of the most common contexts for using mobile devices.
We envision that the personal context is a promising direction for applying mobile AR techniques.
Compared with traditional charts,
ARGVis can be aesthetically pleasing and easy to understand
with the advantage of providing context information directly.
Such characteristics help users comprehend and recall information~\cite{isotype, usefulejunk2}.
ARKit~\cite{ARKit} and ARCore~\cite{ARCore} have been released recently, 
targeting the iOS and Android platforms, respectively. 
These infrastructures provide great opportunities
to utilize mobile AR techniques for visualization in the personal context.
We take the first step toward exploring the application of mobile AR to visualize data for the masses.

\vspace{-2mm}
\subsection{Personal Visualization}
PV focuses on visualizing data in a personal context.
We mainly focus on the mobile PV systems.
Additionally, as some AR content is displayed in the form of real objects,
we also survey physical PV systems, which present data using real objects.
Mobile PV systems visualize data on mobile devices (\emph{e.g.}, mobile phone and tablet),
presenting visualizations to users in a personal context ubiquitously.
Researchers have presented mobile PV systems~\cite{PlayfulBottle, PowerAdvisor, InAir}
to visualize daily information collected from peripheral sensors.
Such mobile systems represent data using standard charts or metaphorical visualizations. 
The former is tedious,  
whereas the latter (whose metaphors are defined by the designers)
may be obscure for general users.
Compared with mobile systems,
physical PV systems are innovative in their data presentation forms, 
which provide abstract data with a concrete physical form~\cite{physicalization-survey}.
Existing works~\cite{evaluating-physicalvis, physical-channels} have investigated physical visualization 
and provided empirical evidence that 
it can improve users' efficiency in information retrieval tasks.
Some researchers have designed physical visualization to visualize personal data.
Khot et al.~\cite{SweatAtoms} designed SweatAtoms, a system that transforms heart rate data into five kinds of 3D printed material artifacts. 
Stusak et al.~\cite{ActivitySculptures} designed four types of sculptures to visualize running activities
and evaluated their usability through a field study.
Their findings confirmed that a physical PV could positively change personal and social behaviors.
However, the physical form of a visualization
is difficult to craft. 
In our tool, users can create visualizations in the form of physical objects based on AR techniques.

\vspace{-2mm}
\subsection{Visualization Toolkits}
Researchers have developed various authoring tools to assist authors in creating visualizations.
Compared to programming toolkits~\cite{d3, vega-lite, web-based-ia}
and WIMP UI systems (\emph{e.g.}, Lyra~\cite{lyra}, iVisDesigner~\cite{ivisdesigner}, and Data-Driven Guide~\cite{DDG})
for information visualization,
pen and touch enable a natural user interface to leverage the human directed manipulation skills,
having
received attention from visualization researchers in recent years~\cite{sketch-story, ivolver, OOD, DOD}.
SketchStory~\cite{sketch-story} 
can quickly create data-bound visualizations on a digital white-board for improvisational purposes.
Mendez et al.~\cite{ivolver} designed and implemented iVoLVER, 
which could create formal visualizations by adopting an interactive visual language.
The most relevant work is DataInk~\cite{DOD}, proposed by Xia et al.,
an authoring tool which supports rigorous creations of data-driven glyph-based visualizations
through direct pen and touch input. We adapt DataInk's approaches on touch interaction to the AR scenario.
Although these systems investigated methods to create data visualizations with pen and touch input,
they mainly focus on the 2D visualization on a 2D canvas,
while AR environments involve 3D.
On the other hand,
while several touch-~\cite{ ARSceneAssembly} 
or pen-based~\cite{SketchingUpTheWorld, MirrorMirror} 
authoring tools have been developed to help users create general AR content,
they do not support the creation of data-driven visualization in AR environments.
Most recently, Sicat et al.~\cite{DXR} presented DXR,
a toolkit which provides both programmatic and graphical user interfaces 
to help designers create prototypes of immersive visualizations rapidly.
However, DXR adopts a form-filling paradigm in the GUI which may not work well in touchscreen devices~\cite{OOD}.
Besides, DXR cannot extract the information from reality to facilitate the creations.
We draw on this line of works to design a tool
that enables non-experts to author data visualization in mobile AR.

\section{Design Rationale}
Given that past work has rarely studied ARGVis,
we try to identify the specificity of authoring ARGVis
through a comparison between 
the abstract workflows of
traditional visualization authoring tools
and
ARGVis authoring tools.
The differences between these two kinds of tools
require us
to characterize ARGVis 
to inform the design of authoring tools.
Thus we shape the ARGVis based on the literature
and finally propose our design considerations.
 
\subsection{The Specificity of Authoring ARGVis}
\label{sec:spec-authoring}

To identify the specificity and challenges in creating ARGVis, 
we compare the abstract workflows of 
two kinds of authoring tools.

\begin{figure}[h!]
	\centering 
	\vspace{-2mm}
	\includegraphics[width=\columnwidth]{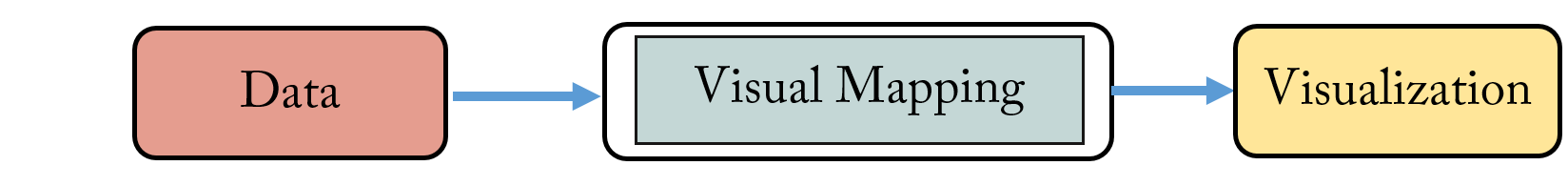}
	\vspace{-8mm}
	\caption{Traditional visualization authoring tools mainly focuses on relationships between data and visual channels.}
	\vspace{-1mm}
	\label{fig:wf1}
\end{figure}

\textbf{Traditional Visualization Authoring Tools.}
One of the core processes of information visualization is 
mapping data dimensions to visual channels (\emph{e.g.}, size, color, position)~\cite{automating-the-design}.
The primary workflow of traditional authoring tools is depicted in~\autoref{fig:wf1},
wherein the input is data and the output is the visualization.

\begin{figure}[h!]
	\centering 
	\vspace{-2mm}
	\includegraphics[width=\columnwidth]{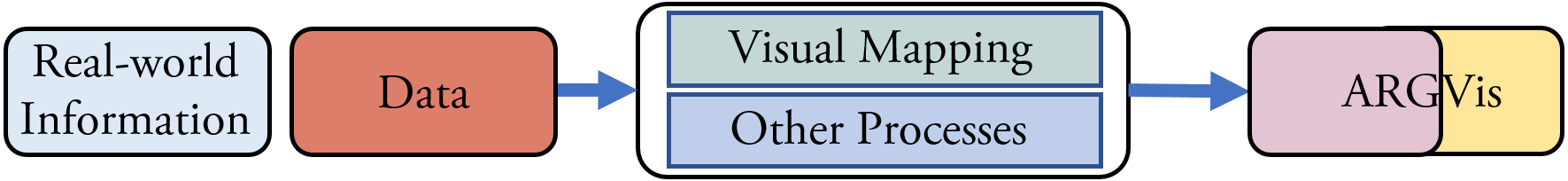}
	\vspace{-8mm}
	\caption{ARGVis authoring tools have various input and output, requiring additional processes to handle relationships between virtuality and reality.}
	\vspace{-1mm}
	\label{fig:wf3}
\end{figure}

\textbf{ARGVis Authoring Tools.}
Compared to traditional visualization authoring tools,
ARGVis authoring tools (shown in~\autoref{fig:wf3}) have two different features:
1) additional input,
which is information about the real-world environment,
and 2) a different output, 
which is ARGVis.
These differences raise 
one critical question for designing an ARGVis authoring tool:
What are the characteristics of ARGVis?

\subsection{The Characteristics of ARGVis}
\label{sec:define-ar-vis}

\begin{figure}[h!]
\centering 
\vspace{-4mm}
\includegraphics[width=\columnwidth]{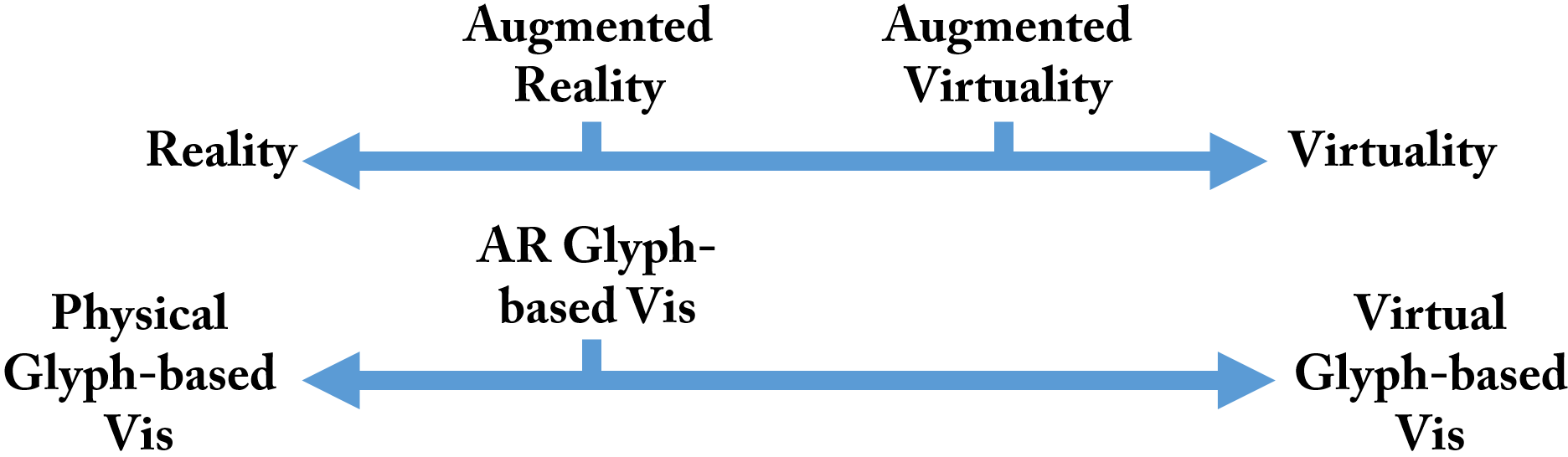}
\vspace{-6mm}
\caption{The glyph-based Vis version of the reality-virtuality continuum.}
\vspace{-2mm}
\label{fig:glyph-continuums}
\end{figure}

Sorting out the characteristics of ARGVis 
can help us identify the specificity of authoring ARGVis 
and form our design considerations.
However, to the best of our knowledge,
there is no existing work that investigates ARGVis.
Hence, we attempt to characterize ARGVis
by considering and summarizing a broad range of 
previous works in visualization.

We first learn lessons from studies of \textbf{AR Visualization}.
A few works exist that have investigated visualizing information in AR. 
Bach et al.~\cite{AR-Canvas} proposed AR-CANVAS 
that depicted
the preliminary design space of AR visualization.
Willett et al.~\cite{embedded-visualization} 
illustrated \emph{situated} and \emph{embedded visualization}.
Although these works focus on general AR visualization,
they inform us that ARGVis has two essential components,
namely, reality and virtuality.

To recognize the related works as much as possible,
we further situate the ARGVis in the reality-virtuality continuum~\cite{ar-continuum} 
(\autoref{fig:glyph-continuums}),
which encompasses all possible variations and compositions of reality and virtuality. 
Focusing on the context of glyph-based visualization,
we recognize 
the left extreme of the continuum,
which consists solely of reality,
corresponds to \textbf{Physicalization}~\cite{physicalization-survey},
which visualizes data using physical glyphs.
Meanwhile the right extreme,
which consists solely of virtuality,
corresponds to \textbf{Glyph-based Visualization}~\cite{glyph-survey},
that visualizes data using virtual glyphs.

Besides,
we note that 
although ARGVis is close to and shares commonalities with Physicalization
(\emph{e.g.}, using 3D real objects as visual marks and bringing data into the real-world),
they have an inherent difference: 
ARGVis is not entirely tangible, 
especially the data part,
making it more like a physicalization without a tangible body.
This characteristic denotes that
ARGVis is also relevant to the concept of 
\textbf{Using Concrete Scales}~\cite{using-concrete-scale} 
which visualizes data with familiar real objects 
but does not rely on their tangible body.

By investigating the reality and virtuality,
and the relationships between them
in
AR Visualization,
Physicalization,
Glyph-based Visualization,
and Using Concrete Scales,
we conclude two key components of ARGVis as follows:
\begin{itemize}
    \item \textbf{Virtuality} consists of virtual glyphs that represent data points using visual channels to encode data attributes~\cite{glyph-survey, physicalization-survey};
    \item \textbf{Reality} is the real-world environment including physical objects~\cite{AR-Canvas, embedded-visualization};
\end{itemize}

\begin{figure}[t!]
	\centering 
    \vspace{-3mm}
    \includegraphics[width=\columnwidth]{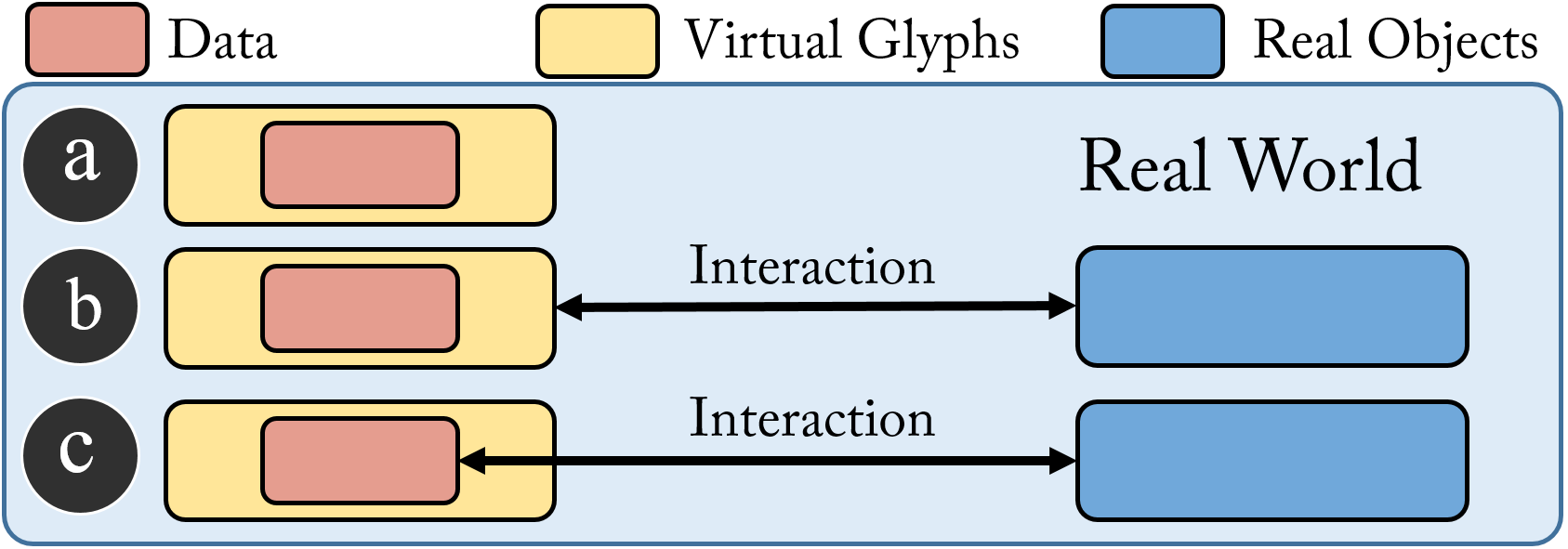}
	\vspace{-8mm}
	\caption{The relationship between reality and virtuality can be a) weak, b) medium, and c) strong.
	}
    \label{fig:connection}
    \vspace{-6mm}
\end{figure}

We summarize three relationships between them 
and describe the rationales and challenges in creating ARGVis to reveal these relationships
as follows:

\begin{itemize}[wide=\parindent]
	\item 
	In the \textbf{Weak Relationship}, the reality only provides a situation 
	background and does not interact with the virtuality directly (\autoref{fig:connection}a).
	In this scenario, 
	ARGVis can bring data into the real world~\cite{physicalization-survey},
	and provides a sense of presence~\cite{cost-benefit-ve}, 
	increasing users' situational awareness.
	To create ARGVis in the \emph{weak relationship},
	users mainly focus on mapping data attributes to visual channels.
	However, the AR environment~\cite{AR-Canvas}
	is more complex and dynamic,
	thereby requiring users to be more careful about visual mappings 
	than in traditional authoring tools.

	\item 
	In the \textbf{Medium Relationship}, real objects, which serve as references, 
	such as the \emph{container} or the \emph{anchor}, are visually related to the 
	3D virtual glyphs (\autoref{fig:connection}b).
	Chevalier et al.~\cite{using-concrete-scale} proposed the concept of \emph{using concrete scales}
	as the process of visually relating complex measures (\emph{e.g.}, extreme magnitudes or unfamiliar units) 
	with familiar objects from the real world.
	Their work inspires us to propose the \emph{medium relationship}. 
	Typically,
	a real object is used as a measure along one of its properties.
	The real object is interpreted for symbolic meaning when measuring along its non-visible properties
	or for direct comparison when measuring along its visible properties (\emph{i.e.,} visual channels).
	When using a visual channel 
	of a real object for comparison,
	the corresponding channel of virtual glyphs must be on the same scale.
	Otherwise, the comparison is meaningless.

	\item 
	In the \textbf{Strong Relationship}, real objects are associated with 
	the data (represented by 3D glyphs) as physical referents~\cite{AR-Canvas, embedded-visualization} (\autoref{fig:connection}c).
	Willett et al.~\cite{embedded-visualization} defined the visualizations in
	this scenario as \emph{embedded visualization},
	one characteristic is being the
	\emph{``positioning individual presentations of data close to their corresponding physical referents.''}
	In the \emph{strong relationship},
	it is tedious for users to manually lay out virtual glyphs close to their physical referents one by one.
	Furthermore, this error-prone process may lead to problematic visualization
	when the positions of virtual glyphs are inaccurate.
\end{itemize}

Our summarization 
can serve as a profile 
that highlights the roles of virtual and real parts 
in ARGVis,
directing our attention 
to design the authoring tool.

\begin{figure*}[b]
	\centering 
	\vspace{-5mm}
	\includegraphics[width=2\columnwidth]{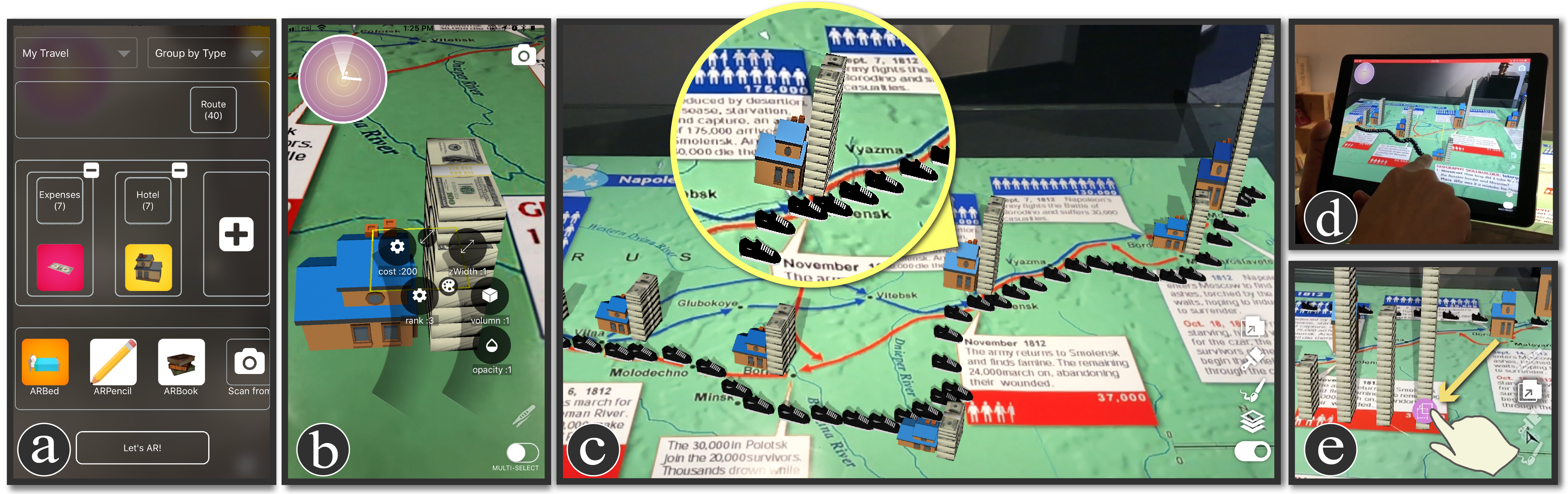}
	\vspace{-3mm}
	\caption{
		a) The UI for mapping data points to virtual glyphs.
		b) The UI for visual mapping. The height of hotels and money stacks represents the cost. 
		The color of houses denotes the rank of hotels. 
		Users can further switch different interaction modes using the buttons in the bottom-right corner.
		c) The ARGVis blends the designer's routes (shoes), hotels (houses), and daily expenses (money stacks) data with a real-world infographic map.
		d) The \emph{Layout Sketch} allows users to lay out numerous objects efficiently. 
		e) The \emph{Copy Layout} method objectifies layouts of collections as small beads. 
	}
	\label{fig:teaser}
	\vspace{-6mm}
\end{figure*}

\subsection{Design Considerations}

We adopt an iterative strategy to formulate our design considerations to guide the design.
These design considerations,
which are rooted in targeted users (non-experts) and the domain (mobile AR),
are first initialized
based on the pipeline of \emph{InfoVis}~\cite{glyph-survey, infovis-pipeline, constructingVis},
relevant authoring tools~\cite{ivisdesigner, DOD, lyra},
and our analysis of authoring visualization in mobile AR (\Cref{sec:define-ar-vis}),
then refined 
through our experiences across multiple design iterations. 

\textbf{DC1. Balance the Expressivity and Simplicity.}
An authoring tool should allow users to customize most 
of the design for expressiveness.
Nevertheless,
the tool should also lower the barrier for designing 
as average users usually do not have enough skill or time to fine-tune their designs~\cite{PV&PVA}.
This requires the tool to make a tradeoff between the expressivity,
which may require comprehensive functionalities,
and simplicity,
which may reduce the expressivity but lower the barrier for non-experts.

\textbf{DC2. Aid Visual Mapping with Context Information.}
In a weak relationship,
the real-world canvas requires users to be more careful about visual mapping 
than in traditional visualization authoring tools~\cite{vis-by-demonstration, voyager},
because it may bias users' visual perception of the visualization~\cite{infovis-perception}.
It is difficult for the users,
especially for those who lack sufficient knowledge of visualization,
to manually adjust the mapping
considering both the data and the real-world contexts.
The tool should leverage context information
to assist users in visual mapping to avoid the pitfalls.

\textbf{DC3. Enable Visual Scale Synchronization between Real Objects and Virtual Glyphs.}
In a medium relationship,
when using a visual channel
of a real object for comparison,
the corresponding channel of virtual glyphs must be on the same scale.
The tool should help users to efficiently synchronize the
visual scales between 
reality and virtuality
for a meaningful comparison.

\textbf{DC4. Support Auto-layout of Virtual Glyphs.}
In a strong relationship,
it is tedious and error-prone
for users to manually place virtual glyphs close to their physical referents one by one.
An ARGVis authoring tool should 
help users map data points with their physical
referents and automatically place them close together.

\vspace{-7mm}
\section{MARVisT}
\label{sec:design}

We present MARVisT, a mobile tool that incorporates our design considerations.
We first describe the general workflow utilizing a usage scenario. 
Then, we introduce the design details of each part.

\begin{figure*}[b!]
	\centering 
	\vspace{-2mm}
	\includegraphics[width=2\columnwidth]{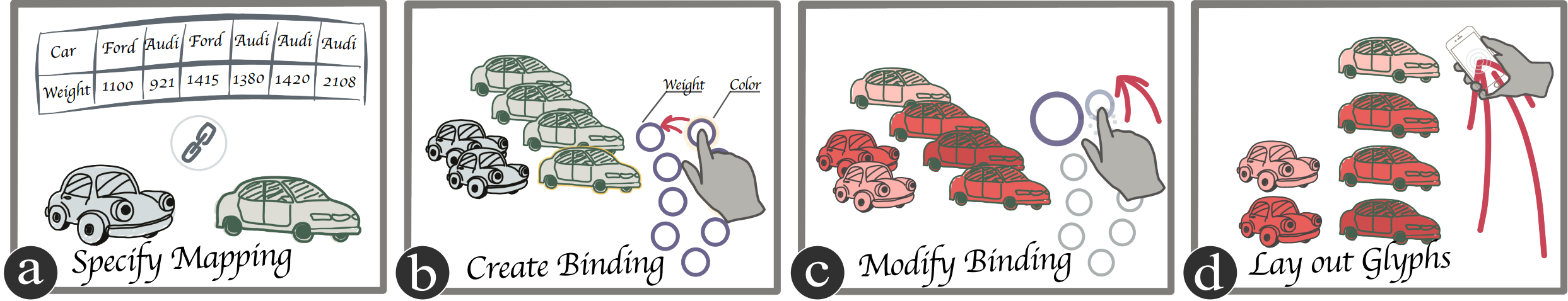}
	\vspace{-2mm}
	\caption{A pipeline based on a bottom-up strategy~\cite{constructingVis} to create glyph-based visualizations.
		a) Specifying the mapping from data points to virtual glyphs.
		b) Creating visual mappings through direct manipulations.
		c) Modifying the scale of visual mappings through one hand operations.
		d) Laying out virtual glyphs through intuitive and effective interactions. }
	\label{fig:pipeline}
	\vspace{-6mm}
\end{figure*}

\vspace{-2mm}
\subsection{Usage scenario}


Suppose Bob 
goes to a museum on the last day of his journey.
During his visit,
an infographic map depicting the path of Napoleon's march to Moscow attracts Bob's attention.
Bob wants to visualize his travel route in Russia 
on the map for comparison to get a sense of history.
His mobile phone has recorded various travel data,
including routes (\emph{e.g.}, distance, and duration of travel), 
hotels (\emph{e.g.}, cost and rank), 
and daily expenses (\emph{e.g.}, cost).
Bob exports the 
data into MARVisT.
In the visual mapping interface (\autoref{fig:teaser}a), 
he chooses the shoe, the house, and the money stack models to 
represent the route, hotel, and daily expense data, respectively.
MARVisT automatically duplicates the models for each data point,
distributing them methodically on the map.

To encode the \emph{distance} of his routes,
he taps on a shoe to open the palette for creating the visual mapping.
The palette 
looks like
a double-ring semi-annulus (\autoref{fig:teaser}b), 
where the inner ring consists of beads representing data attributes 
and the outer ring displays the visual channels.
Currently, MARVisT supports a suite of frequently-used visual channels, 
such as \emph{size (1D-length, 2D-area, 3D-volume), color, angle}, and \emph{opacity}.
When Bob taps the \emph{distance} bead,
MARVisT recommends the \emph{x-width} channels 
by highlighting the 
bead based on Bob's current point-of-view 
and the ordered data type of \emph{distance}.
Following the suggestion,
Bob creates the mapping by dragging the \emph{x-width} bead 
and dropping it to the \emph{distance} bead.
He modifies the value of \emph{x-width} so that the shoe model has the same scale as the map.
Then, Bob 
leverages the \emph{y-height} channel of the house models to encode the \emph{cost} attribute.
MARVisT propagates the same kind of mapping to all data points that have the same attribute.
Hence all the money stack models, 
which 
also have the \emph{cost} attribute, 
are likewise encoded with the \emph{y-height} channel (\autoref{fig:teaser}b).
Next, 
Bob decides to encode the \emph{rank} attribute of houses using the \emph{z-length} channel.
As the \emph{y-height} channel of houses  
has already been used for visualizing the \emph{cost} attribute,
which may lead to interference with the \emph{z-length} channel,
MARVisT sends a warning message to notify Bob of this pitfall.
Thus, he uses the \emph{color} channel to encode the \emph{rank} instead.

MARVisT enables users to lay out the glyphs through several methods.
By dragging the houses to their cities one-by-one,
Bob visualizes the hotel information on the map.
Then, Bob 
groups the data points into three collections based on model types
and
invokes the \emph{Copy Layout} function to transfer the house layouts
to the money stack collection (\autoref{fig:teaser}e),
thus distributing the money stacks alongside the corresponding houses.
Now he can easily compare hotel cost and expenses every day.
It is time to portray the itinerary on the map.
Although it is possible to recognize a map and place virtual models based on GPS data automatically,
the infographic map is customized,
leading to difficulty in recognition and unpredictable mismatches between the route data and the map.
Therefore, Bob needs to place these virtual glyphs by himself.
By using the \emph{Layout Sketch} function,
he sketches a path along with the house and money stacks on the screen
to distribute the shoes on the map (\autoref{fig:teaser}d),
revealing his travel route.
Comparing his routes to Napoleon's,
he discovers that his path is similar to that of Napoleon's retreat from Moscow (\autoref{fig:teaser}c).
``Uh-huh! Such a fading empire'', he muses.
Bob 
captures this ARGVis as an image
and sends it to his friends.

\vspace{-2mm}
\subsection{Basic Pipeline}
\label{sec:basic-pipeline}

To incorporate DC1,
MARVisT adopts 
a \emph{bottom-up strategy}~\cite{constructingVis, glyph-survey} (\autoref{fig:pipeline})
to guarantee the expressivity
and an \emph{objectifying strategy}~\cite{OOD, DOD}
to ensure the simplicity of interactions.

\textbf{Map Data Points to Visual Marks.}
MARVisT currently only supports data in tabular format.
After importing data,
users can drag different groups of data points 
from the data gallery
and 3D glyphs from the glyph gallery 
to the middle slots (\autoref{fig:teaser}a) 
to specify the mapping between them.
Currently, MARVisT only supports model fetching from servers,
since using advanced methods,
such as reconstruction from real objects using a depth camera~\cite{kinectfusion},
to acquire 3D models is out of the scope of this research.

\textbf{Create, Modify, and Remove Visual Mappings.}
While the form-filling paradigm acts as the core of WIMP UI,
such designs are tedious, slow, and error prone in touchscreen devices~\cite{OOD}.
Inspired by DataInk~\cite{DOD},
we objectify data attributes and visual channels as interactive beads
that can be directly manipulated to get rid of form-filling paradigm.
Users can create a mapping through a drag-and-drop interaction (\autoref{fig:pipeline}b).
Then, MARVisT will automatically propagate the mapping
to the corresponding data attribute and visual channel of all virtual glyphs
based on their data values.
To modify the scale of a bound visual channel,
users can activate the modification state and then slightly drag the bead up or down to alert the value (\autoref{fig:pipeline}c).
The modification will again be propagated to all glyphs.
For deletions, users only need to drag the visual channel away from the paired data attribute.
The motion will clean the mapping from all the glyphs.

\textbf{Lay Out the Glyphs.}
After the visual mapping,
users can further author the layout of virtual glyphs.
MARVisT follows the suggestion of previous works~\cite{3dmanipulation-survey}
to avoid cross-dimensional interactions,
supporting four kinds of interactions
, namely, $(2D, 3D) \times (Glyph, Collection)$,
to lay out the virtual glyphs.

\begin{figure}[h]
\centering 
\vspace{-2mm}
\includegraphics[width=\columnwidth]{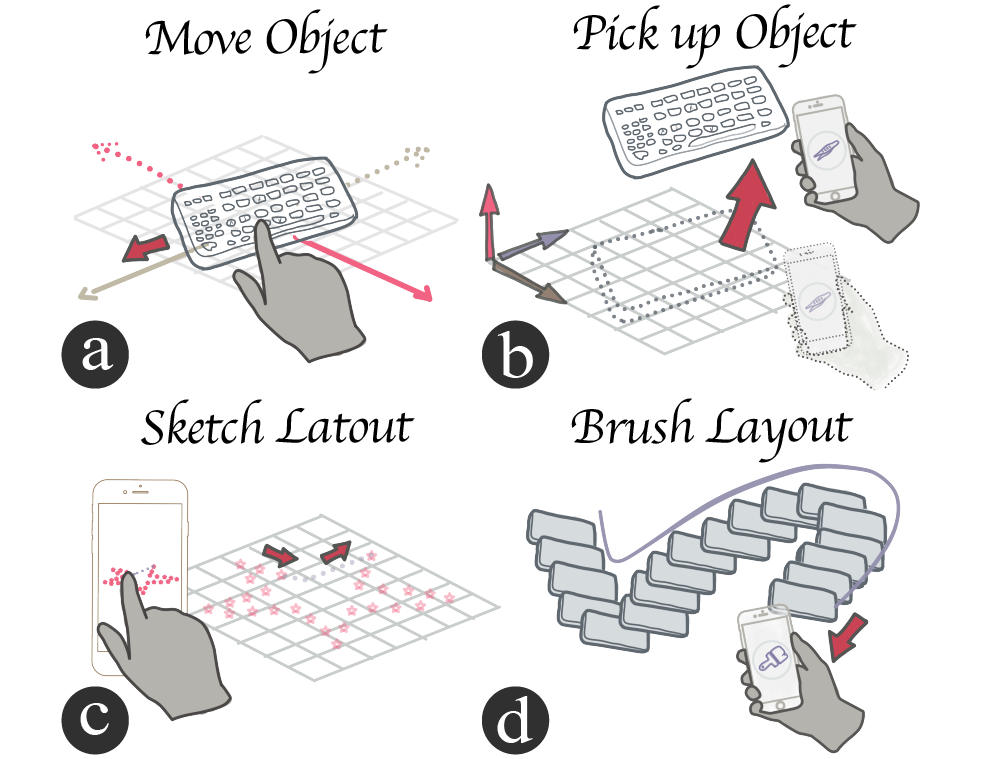}
\vspace{-4mm}
\caption{
	MARVisT provides intuitive interactions that allow users to lay out single or multiple virtual glyphs in both 2D and 3D space efficiently.}
\vspace{-5mm}
\label{fig:interaction}

\end{figure}

\begin{itemize}[leftmargin=*]
	\item \underline{\emph{2D Interactions on Glyph Level.}}
	MARVisT enables users to directly manipulate the position of virtual glyphs using familiar gestures (\autoref{fig:interaction}a).
	For simplification,
	the object is limited to move on the plan it located
	when interacting through touching. 

	\item \underline{\emph{3D Interactions on Glyph Level.}}
	As for the 3D positions of the glyphs,
	we design a \emph{Picker} mode (\autoref{fig:interaction}b),
	wherein users can use the mobile as tongs to grip and lift glyphs in the 3D space.

	\item \underline{\emph{2D Interactions on Collection Level.}}
	MARVisT allows users to group glyphs into collections based on data attributes.
	A collection is treated as a single object of a higher level,
	to which the 
	interactions above can also be applied.
	Inspired from DataInk~\cite{DOD},
	we design a \emph{Layout Sketch} method 
	through which users can draw a path on the screen (\autoref{fig:interaction}c) 
	to distribute glyphs in the selected collection along the path.
	This method will only modify the horizontal positions of glyphs.

	\item \underline{\emph{3D Interactions on Collection Level.}}
	To modify the 3D position of a collection,
	users can also use the \emph{Picker} mode to position a collection anywhere in the 3D space.
	We further design a \emph{Layout Brush} to enable the laying out of glyphs in the 3D space (\autoref{fig:interaction}d).
	Users can 
	wave their mobile like a physical brush in the air to distribute the virtual glyphs in the selected collection
	to the real-world 3D space.

\end{itemize}

To facilitate the creation and save users from tedious work,
we design several smart functions
for efficiently laying out glyphs for data visualization purposes,
which can be extended in the future:

\begin{itemize}[leftmargin=*]
	\item \underline{\emph{Side-by-side Comparisons.}}
	By triggering the \emph{Copy Layout} method,
	users can just drag the layout of the source collection and drop it to the target collection
	to transfer the desired layout.
	
	\item \underline{\emph{Stacking Glyphs.}}
	MARVisT will detect the positions of glyphs in a collection 
	whenever its layout is modified
	and help users to align glyphs automatically
	for stacking glyphs (\autoref{fig:pipeline}d).
\end{itemize}

\subsection{Context-aware Nudging}

The complex and dynamic AR environment~\cite{AR-Canvas}
brings difficulties for visual mappings.
Considering both best practices guideline drawn from prior research~\cite{automating-the-design, infovis-perception, tamara-visualization}
and the context information,
MARVisT validates the usage of visual channels
before and after creating a visual mapping
and guides users with the validation results through an indirect way (DC2).


\textbf{Validation.}
From the perspective of perceptual effectiveness,
MARVisT considers the following features in the validation to cover all supported visual channels: 

\begin{itemize}[leftmargin=*]
\item \underline{\emph{Orientation.}}
The size channels (\emph{e.g.}, \emph{1D-length}, \emph{2D-area}, and \emph{3D-volume})
are widely used for ordered data.
In the AR environment, the perspective distortion, 
an important depth cue of the 3D environment,
strongly interferes with these size channels~\cite{infovis-perception}.
Thus an intuitive assumption is that 
if a size channel has components along the depth axis,
it will be ineffective.
MARVisT calculates the depth component of each size channel based on users' perceiving \emph{orientation}
and invalidates the one which has the depth component.
One limitation of this heuristic rule is that 
in mobile AR environments the authoring perspective might not be the viewing one.
Nevertheless,
this rule is still effective
as the authoring perspective is usually the best angle for viewing
and the user can be informed of the pitfalls of using depth channels.

\item \underline{\emph{Luminance Contrast.}}
Whenever users encode data using optical channels (\emph{e.g.}, \emph{color} and \emph{opacity}),
MARVisT validates the \emph{luminance contrast} between the virtual glyphs and the real-world background.
Adequate \emph{luminance contrast} should be ensured 
for perceiving shape, form, and other details~\cite{infovis-perception}.
MARVisT adopts the root-mean-squared (RMS) method~\cite{local-luminance}
to calculate the contrast ratio given that the AR environment is closer to natural images.
A minimum contrast ratio between a pattern and its background is suggested to be 3:1~\cite{infovis-perception}.
MARVisT will fail the validation 
if the visual encoding cannot meet the minimum contrast ratio.
While our rule aims to ensure the legibility,
the color issue in the AR environment is far more complex and involves many 
physical and physiological 
factors.
We further discuss color issues in~\Cref{sec:discussion} and leave them for future work.

\item \underline{\emph{Rotational Symmetry.}}
For the \emph{angle} channels, MARVisT will validate the \emph{rotational symmetry} of the 3D glyph.
MARVisT will fail the validation 
if the 3D glyph's order of \emph{rotational symmetry} about the axis is greater than four (means that the rotation by 90$^{\circ}$ does not change the object)
since the perceptual accuracy of angles drops off within the exact horizontal and vertical~\cite{infovis-perception}.

\item \underline{\emph{Separability.}}
MARVisT also considers that the visual channels have already been used 
to avoid subsumption and perceptual integration~\cite{infovis-perception}.
For example, when encoding data attribute $a_i$,
if the \emph{x-width} channel is used for attribute $a_j$,
\emph{area} channels will be invalided to avoid subsumption;
\emph{y-height} channel will also be invalided to maintain the separability of the channels.
\end{itemize}

\begin{table}[h!]
	\vspace{-4mm}
	\caption{Permitted Visual Channels of Data Types.}
	\vspace{-4mm}
	\label{table:2}
	\begin{threeparttable}
	\centering
		\begin{tabular}{l l}
			\toprule
			\textbf{Data Types} & \textbf{Visual Channels\tnote{*}} \\ \midrule
			\textbf{Quantitative} & length (x-, y-, z-) $>$ angle ($\varphi$-, $\theta$-, $\psi$-) \\
			& $>$ area (top-, left-, front-) \\ 
			& $>$ color (luminance or saturation) $>$ volume \\ \midrule
			\textbf{Ordinal} & color (luminance or saturation) \\ 
			& $>$ length (x-, y-, z-) $>$ angle ($\varphi$-, $\theta$-, $\psi$-)  \\
			& $>$ area (top-, left-, front-) $>$ volume \\ \midrule
			\textbf{Nominal} & color (hue) \\ \bottomrule
		\end{tabular} 
		\begin{tablenotes}\footnotesize
			\item[*] Ordered by perceptual effectiveness rankings.
		\end{tablenotes}
	\end{threeparttable}
	\vspace{-2mm}
\end{table}

\textbf{Nudging.}
Some existing works~\cite{voyager} 
automatically generate the visualization encoded with the most effective visual channels
and prohibit any ineffective combinations of them.
However,
targeting the masses produces the requirements of
enabling flexible and expressive creations for greater engagement,
which may violate design principles of visualization.
Considering these subtle requirements of general users,
we prefer to nudge users instead of prohibiting them with hard constraints.
Nudging is a concept that proposes positive reinforcement and indirect suggestions to influence users~\cite{nudge}.
MARVisT integrates a twofold nudging mechanism, meaning it, 
nudges users in 1) what to do and 2)  what not to do.
Specifically,
when the user is ready to encode a data attribute with visual channels,
MARVisT will generate a list of valid visual channels ordered by perceptual effectiveness~\cite{tamara-visualization, infovis-perception, automating-the-design} in relation to the data type (\autoref{table:2}).
It would then
recommend the most effective one to the user through a jumping animation of the bead.
When the user finishes creating or modifying a visual encoding,
MARVisT will validate the visual encoding
and warn the user through messages if the validation fails.

Admittedly, our initial design is modest. 
Although more advanced context-aware nudging methods are possible, 
the primary goal of this proof-of-concept work is 
to explore and design features of an overall approach to enable non-experts 
to create ARGVis. 
In place of 
more 
advanced
methods, 
we intentionally 
use interpretable, deterministic heuristics for nudging users with context awareness. 
We envision the design and evaluation of improved context-aware nudging systems
as an important future research direction.

\vspace{-4mm}
\subsection{Visual Scales Synchronization}
\label{sec:virtualvisualchannels2realvisualchannels}

\begin{figure}[h]
\centering 
\vspace{-3mm}
\includegraphics[width=\columnwidth]{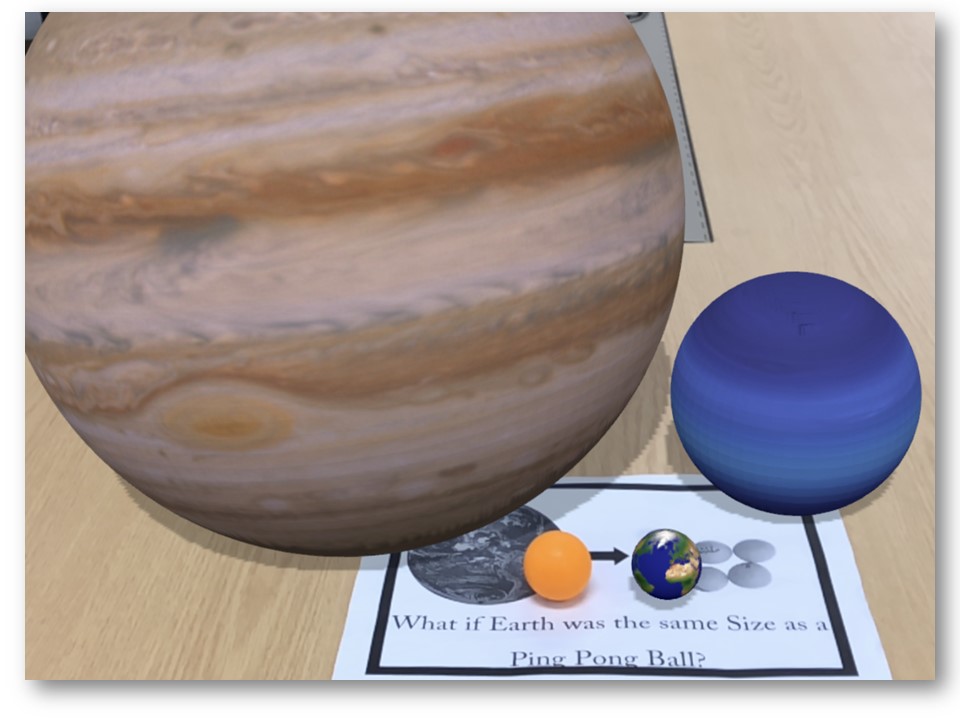}
\vspace{-9mm}
\caption{
Users can extract the \emph{volume} of the ping pong ball
and assign it to the Earth. 
Other virtual planets (Neptunus and Jupiter) will be scaled accordingly 
if the corresponding visual mapping on the \emph{volume} exists.}
\label{fig:earth}
\vspace{-3mm}
\end{figure}

As described in DC3,
if a visual channel of real objects (referred to as \emph{RVC}) is used for comparison,
the corresponding visual channel of virtual glyphs (referred to as \emph{VVC})
must be on the same scale as the \emph{RVC}.
For example,
if a designer wants to 
compare the \emph{volume} channel of Earth, Neptune, and Jupiter 
using a ping pong ball as a benchmark
 for familiarity and intuition,
he needs to change the size of any planet to the same size 
as the ball (\emph{e.g.}, the Earth in \autoref{fig:earth})
and scale other planets accordingly to maintain the relative ratio of their \emph{volume}
(converting their \emph{volume} channels from the original scale to the ping pong ball scale).

Although the user can manually adjust a \emph{VVC} 
to synchronize with an \emph{RVC}
via the basic interactions (\Cref{sec:basic-pipeline}),
it is tedious and ineffective.
We design a semi-automatic function based on object detection techniques~\cite{Vision, ARKit}
to improve the effectiveness.
After detecting a real object,
MARVisT automatically extracts the values of its visual channels,
such as \emph{size, position, text}, and so on.
Limited by the computer vision technique,
currently,
MARVisT only supports an approximation of these values.
For instance, we estimate a real object's size channels 
based on its bounding box.
Also, MARVisT can only detect text on a pure color background.

When the user taps on this real object,
a single-ring semi-annulus similar to the one in \autoref{fig:teaser}b
will pop out.
The semi-annulus consists of beads which represent 
the available visual channels of the real object.
With a similar interaction,
the user can drag a bead 
and drop it onto a virtual glyph 
to assign its value to the corresponding visual channel of the virtual glyph.
If this visual channel has already been used to encode a data attribute,
MARVisT will inversely calculate the new scale
based on the value of the \emph{RVC} 
and the data bounded with the virtual glyph.
Then the new scale will 
automatically 
be
propagated to all virtual glyphs with the same visual mapping,
leading to updates of corresponding \emph{VVC}s.
For more technical details on the object detection, 
the visual channel extraction,
and \autoref{fig:earth},
we refer the reader to the appendix.

\subsection{Virtual Glyphs Auto-layout}

\begin{figure}[h]
\centering 
\vspace{-4mm}
\includegraphics[width=\columnwidth]{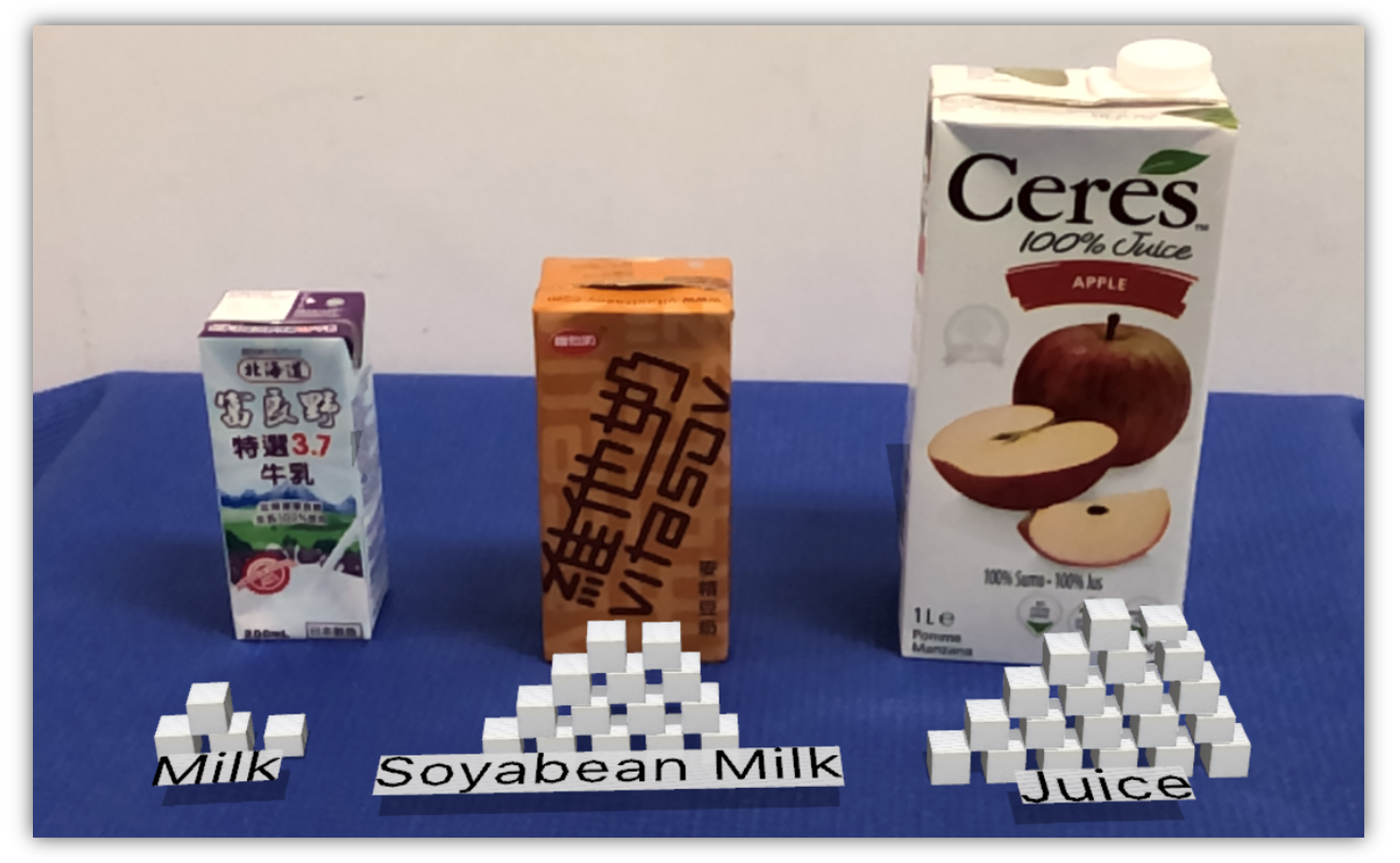}
\vspace{-7mm}
\caption{The sugar content of drinks. 
If there are $n$ virtual sugar stacks needed to be placed,
the user needs to perform $O(n)$ interactions.
With the auto-layout,
the user can finish the task using $O(1)$ interactions.}
\label{fig:sugar}
\vspace{-3mm}
\end{figure}

As demonstrated in DC4,
an authoring tool should help users to automatically place
virtual glyphs (referred to as \emph{VG}s) together 
with their corresponding real objects (referred to as \emph{RO}s).
For example, 
without supporting the automatic laying out process,
a user needs to manually move the three virtual sugar stacks
one-by-one
to create what we see in \autoref{fig:sugar}.

For the auto-layout,
the tool must first 
detect \emph{RO}s,
map data points to the \emph{RO}s,
and then place \emph{VG}s of the data points
to the positions of their corresponding physical referents.
For real object detection,
we can reuse the methods mentioned in \autoref{sec:virtualvisualchannels2realvisualchannels}.
As for the second step,
to enable the one-to-one mapping process fully automatically
without using a hard-code method
is non-trivial,
given the identifiers
of the \emph{RO}s
depending on the underlying recognition algorithm.
For instance, the milk in \autoref{fig:sugar}
can be recognized as different identifiers (\emph{e.g.}, \emph{milk} or \emph{drink}, etc.) 
using different computer vision libraries,
leading to a difficulty in mapping the data points to the \emph{RO}s.
To decouple the identification of \emph{RO}s from the recognition algorithms
and
design a generalizable approach 
for mapping data points to \emph{RO}s,
we propose to use a semi-automatic approach
that asks the designer to
specify the mapping
by using the data attributes of data points
and the visual channels of \emph{RO}s.
We chose visual channels to depict a real object because
1) the visual channel is one of the most accessible features for a mobile phone with a camera,
2) the values of visual channels are independent of the underlying recognition algorithm, 
and 3) it is convenient to reuse visual channels 
that have already been extracted in \Cref{sec:virtualvisualchannels2realvisualchannels}.
Moreover, our approach does not require extra UI widgets and interactions.
Specifically, to specify the mapping, 
the user can drag the bead of 
a visual channel of the \emph{RO}s
and drop it onto the bead of 
a data attribute of the \emph{VG}s.
MARVisT will 
map the \emph{RO}s
to the data points 
by their index 
in the selected channel and attribute.
For example,
the user can drag the bead of
the \emph{volume} channel of the drinks in \autoref{fig:sugar}
and drop it onto the \emph{sugar content} attribute of the data points (presented by sugar stacks)
to map them
by index:
the data point with the highest \emph{sugar content}
to the drink with the highest \emph{volume}, and so on.
Finally,
MARVisT will automatically place each glyph 
to the position of its corresponding real object.
The user can further fine-tune the \emph{VG}s' positions
through the layout interactions introduced in \Cref{sec:basic-pipeline}.

Currently, MARVisT only supports mapping between categorical data 
and ordinal data with monotonic relationships.
Other cases are left for future works.
The technical details and
the step-by-step explanations of creating \autoref{fig:sugar}
are enclosed in the appendix.
%
%


\begin{figure*}[bh!]
\centering 
\vspace{-4mm}
\includegraphics[width=2\columnwidth]{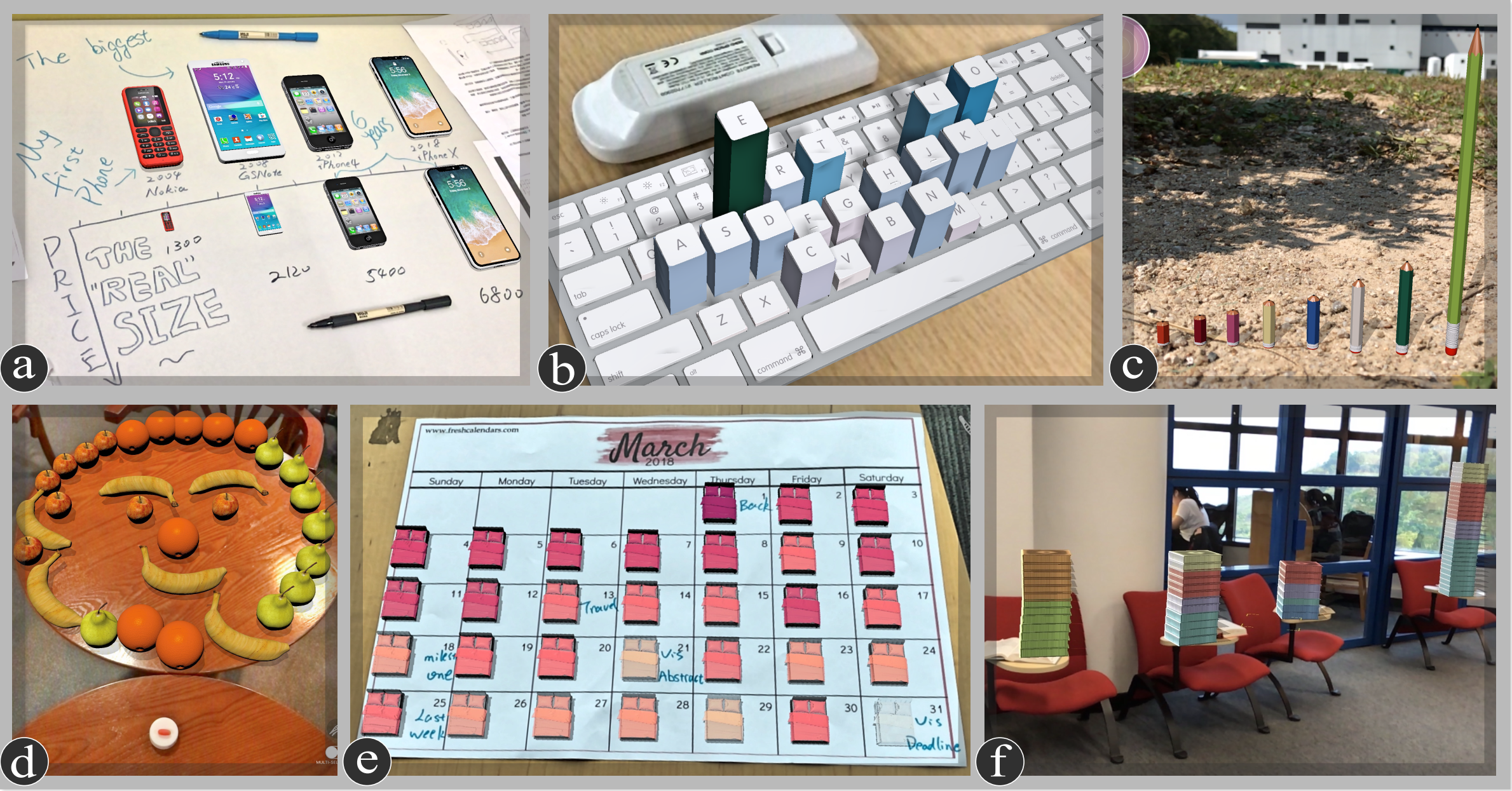}
\vspace{-2mm}
\caption{Examples created with MARVisT. 
	a) An isotype chart combines the reality (the sketching on the whiteboard) 
	and the virtual glyphs (the eight mobile phones). The four mobile phones in the second row visualize their price using the size channel.   
	b) Michael Knuepfel's keyboard frequency sculpture recreated and enhanced using MARVisT. Both the color and height channels are used for double encoding.
	c) Moodley brand identity's physical infographics recreated by MARVisT using color and height to encode various data attributes.
	d) A smiling face created using the \emph{Layout Sketch} method. The face consists of fruits the designer eat in a week. 
	e) A calendar visualization created using the auto-layout method. The design displays the sleep duration per night in the March 2018 using the color and opacity for double encoding.
	f) A stacked glyph chart created using the \emph{Layout Brush} and the automatically stacking function. It depicts the designer's favorite seat in the library. Color represents the day of the week.}
\label{fig:examples}
\vspace{-5mm}
\end{figure*}

\vspace{-4mm}
\section{Implementation and Performance}
MARVisT is a universal mobile application implemented on iOS
with techniques that include React Native~\cite{ReactNative},
ARKit~\cite{ARKit},
Vision~\cite{Vision},
and SceneKit~\cite{SCENEKit}.
The tracking of real-world environments
and registration of virtual glyphs are supported by ARKit,
a powerful infrastructure that enables the development of AR applications.
Also, in each frame of the ARKit session,
we utilize the HSL image of the field-of-view to calculate the luminance contrast ratio 
and the environment light estimation to improve the realism.
To recognize real objects and estimate visual channel values from them,
MARVisT leverages both 
the Vision (wherein the model for object recognition was trained in Turi Create using Darknet YOLO) 
and ARKit.
We embedded a toy dataset of reference objects in MARVisT
to supplement and improve the performance of extracting visual channel values.
The dataset can then be extended and shifted to the cloud in the future.
MARVisT adopts a responsive design that allows adaptation to mobile devices with different screen sizes.
The design of MARVisT does not depend on unique techniques listed above.


We conducted experiments to evaluate the construction times, 
the frame rates of the static scenes and the dynamic scenes 
for varying data sizes running on an iPhone 8 Plus and an iPad Pro
to identify the limits of MARVisT.
The detailed experiments' settings and results are found in Appendix B.
Overall, for a dataset with reasonable size (1, 000 models or less), 
our implementation achieves 
a low construction time (in around 2 seconds)
and real-time frame rates (over 50 FPS in most cases).
Considering the usage scenario (i.e., the personal context) of MARVisT, 
wherein users likely do not have big datasets to visualize, 
we think the performance of the current implementation is acceptable.
We believe 
MARVisT still can be improved
in the future, such as 
optimizing the memory usage and 
following the best practice, \emph{etc}.

\vspace{-3mm}


\section{Evaluation}
\label{sec:userstudies}
MARVisT was evaluated from two perspectives. 
First, we presented a gallery of works created by MARVisT to illustrate 
its expressiveness. 
Then an in-lab user study with non-experts was conducted to evaluate MARVisT's usability and usefulness.

\vspace{-4mm}
\subsection{Example Personal Visualizations}
To demonstrate its expressiveness, 
we created a diverse set of ARGVis using MARVisT. 
MARVisT supports a wide range of glyph-based visualizations 
from one glyph encoded with a single visual channel 
to a collection of glyphs with multiple visual channels.
Moreover, devising engaging real-world infographics using MARVisT is more efficient than manually designing physical visualizations, 
including material selection and object placement, 
which illustrators and designers are usually responsible for.

MARVisT provides several visual channels for encoding data,
which is essential for attaining data-driven design. 
For example, Figure~\ref{fig:examples}a presents a sample visualization
that combines virtual glyphs 
with real-world content 
thereby depicting previously-used mobile phones of the designer
whose sizes visually encode the associated price.
Users can also 
encode various data attributes with different visual channels (\autoref{fig:examples}c),
or use multiple channels to encode a data attribute
for multiple encoding
(\autoref{fig:examples}b, e). 

Besides,
customizations become increasingly flexible 
when users manipulate the layout of virtual glyphs in collections. 
For example, a user can 
distribute the glyphs along a sketch path on a plane to create a customized layout (\autoref{fig:examples}d). 
Figure~\ref{fig:examples}e 
organizes the glyphs in a calendar layout.
MARVisT also supports the laying out of glyphs in 3D space through the \emph{Layout Brush}
methods and enables users to stack glyphs with minimal effort
by guessing the users' brush intentions (\ref{fig:examples}f). 

Utilizing the object detection technique,
MARVisT can reduce users' workload in creating designs.
Similar to the case in \autoref{fig:earth},
MARVisT can extract the 
geometric properties (\emph{e.g.}, top-view size) of real objects
and assign them to
the bars in \autoref{fig:examples}b,
efficiently adjusting their size to make them fit with the real background key-caps.
Additionally,
through building connections between
visual features of real objects (\emph{e.g.}, the \emph{text} of the whiteboard in \autoref{fig:examples}a, key-caps in \autoref{fig:examples}b, and the calendar in \autoref{fig:examples}e) 
and data attributes of virtual glyphs,
MARVisT can automatically place multiple virtual glyphs 
onto their corresponding physical referents,
thereby improving the efficiency.

Adopting the data-driven strategy,
MARVisT dynamically updates the graphics according to changes 
of the 
dataset.
For example, 
if 
the price of mobile phones in~\autoref{fig:examples}a is changed, 
the size of mobiles will be updated accordingly.
Given the data-driven feature and the physical form of virtual glyphs,
it is highly efficient to devise physical infographics with MARVisT.
For instance,
we recreated the physical infographic 
Figure \ref{fig:examples}b from Michael Knuepfel~\cite{MichaelKnuepfel}
and Figure \ref{fig:examples}c from moodley brand identity~\cite{moodleybrandidentity}.
A designer using MARVisT does not need to purchase, 
craft, and place the physical materials
but can still obtain the same visual effects as the traditional physical infographics does.



\subsection{User Study}
Next, we conducted a laboratory study to examine 
if non-experts could create ARGVis using MARVisT
and if there were any usability issues for improving the system.
The study also aimed to gain 
of presenting data in the form of ARGVis.

\vspace{-2mm}
\subsubsection{Participants and Experimental Setup}
We recruited 14 participants (5 males; ages:21 to 27, average: 23.6)
from the university through email announcements and billboard advertisements.
We pre-screened participants to
ensure that they all had no or limited experience in data visualization design.
According to the pre-study survey,
participants had diverse major backgrounds,
including Computer Science (4),
Chemistry (2),
Electronic Engineering (2),
Economics (1),
Social Science (1),
Food Science and Engineering (1),
Plant Protection (1), 
Digital Media (1),
and Big Data Technology (1).
They were also not the co-authors of the paper.
All of them reported 
having prior experience in mobile AR,
but none had created AR content before.
The participants were identified as P1-P14, respectively.
Each participant received a gift card worth \$6.37 
at the beginning of the session,
independent of their performance.
The study was run in a conference room which was $5m \times 4m$,
using a 10.5-inch iPad Pro.

\subsubsection{Procedure}
Participants accomplished a consent form and a demographics questionnaire,
completed three tasks with MARVisT, 
and then concluded with a post-study survey and an interview.

\textbf{Demonstration and Training (25min).}
The investigator demonstrated 
the concept of glyph-based visualization in mobile AR 
and then introduced MARVisT, 
including its functions and interactions
in the glyph mapping, visual mapping, and layout steps 
through several walk-through examples. 
Participants were asked to exactly replicate the provided examples, 
and seek guidance and assistance whenever necessary. 
The participants were also encouraged to freely try and explore 
every function and interaction until they felt confident about using the tool.

\textbf{Replication (25min).}
The investigator proceeded to the replication task.
In this phase, the participants were asked 
to replicate two data visualizations 
(\autoref{fig:teaser}c and \autoref{fig:examples}b, referred to as RT1 and RT2) 
for two new datasets
based on the real-world background (the infographic map and the keyboard).
When the task began, participants received 
documentation about the datasets and
several color images with legends of the targeted visualizations.
We designed the visualizations to be sufficiently rich in features
to enable participants to reach most of the functions in MARVisT during the replication.
The study setup was informal,
allowing participants to interrupt at any time to ask questions.

\textbf{Freeform Exploration (20 min).}
Next,
the investigator asked participants to freely explore the tool by designing their own data visualizations
with pre-loaded datasets.

\textbf{Post-study Survey and Interview (20 min).}
Participants were asked to complete a post-study survey
using a 7-point Likert scale (1-strongly disagree, 7-strongly agree).
The questions assessed the usability (Q1-Q4) and expressiveness (Q5) for DC1,
the usefulness (Q6-Q8) of the features to accommodate DC2-DC3,
and the engagement (Q9) using MARVisT.
The investigator then conducted a semi-structured interview to collect
qualitative comments.

\subsubsection{Results}

\begin{figure}[h]
	\centering 
	\vspace{-4mm}
	\includegraphics[width=1\columnwidth]{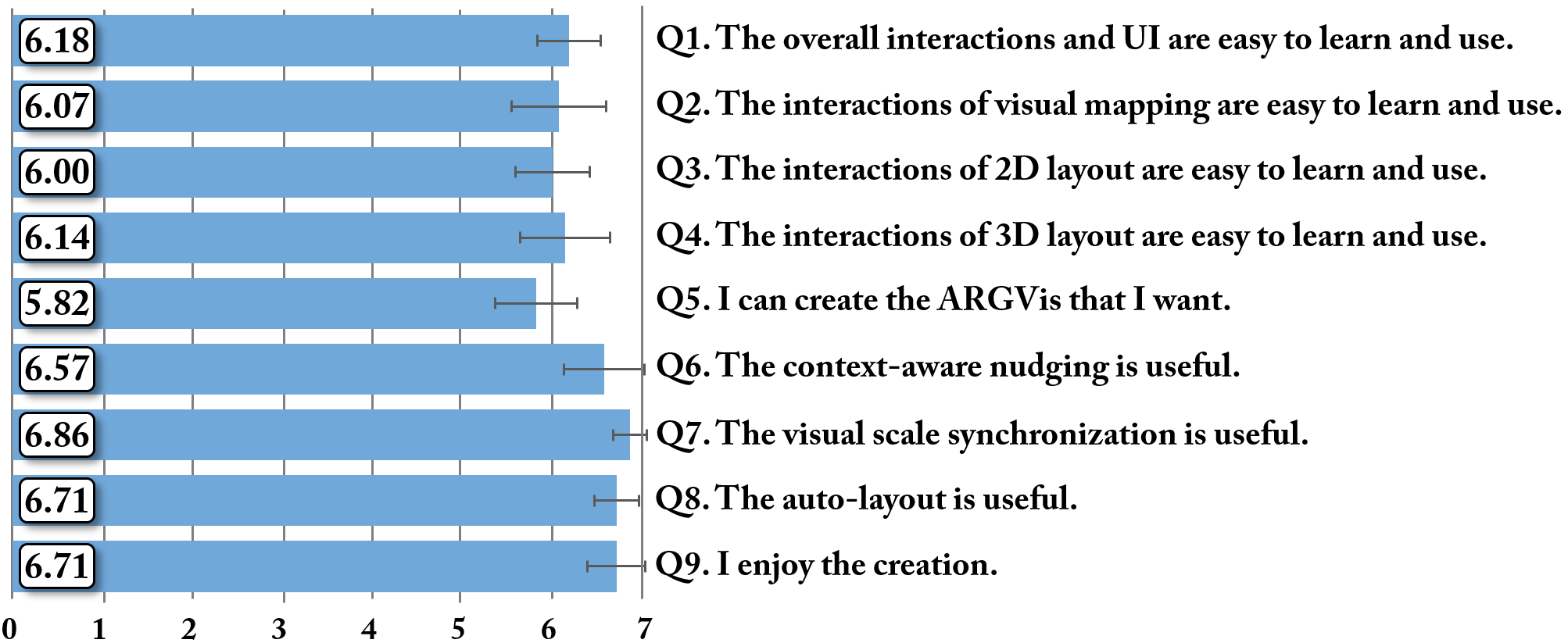}
	\vspace{-8mm}
	\caption{All questions received an averaging rating greater than five, which was very encouraging.}
	\label{fig:questions}
	\vspace{-4mm}
\end{figure}

All the participants completed the replication tasks with minimal guidance
in roughly 13 min in total.
The third task was open-ended 
and involved interview-like conversations 
between participants and the investigator
that derived useful feedback for the tool and ARGVis.
\autoref{fig:questions} presents the questions and average user ratings.
Our analysis of subjective ratings 
and qualitative comments of the participants
led to insights into the usability, expressiveness, usefulness, and engagement of the authoring experience.

\textbf{Usability.}
Targeted at general users, 
MARVisT must be both easy to learn and use.
Overall, the usability of MARVisT was appreciated by the participants
based on the result of Q1 ($\mu=6.18$, 95\% $CI = [5.83, 6.53]$).
Specifically,
based on the results of Q2, Q3, and Q4,
the participants responded positively about the
the main interactions of MARVisT 
(all the ratings of the three questions were not lower than 6). 
They found the objectification UI design 
to be \emph{``effective and reasonable''} (P1, P3, P4, P6-P14)
and rated the interactions for visual mapping 
as easy to learn and use ($\mu=6.07$, 95\% $CI = [5.55, 6.59]$).
The participants also felt it was easy to learn and use 
the 2D layout interactions ($\mu=6.00$, 95\% $CI = [5.59, 6.41]$)
and the metaphorical 3D layout interactions ($\mu=6.14$, 95\% $CI = [5.64, 6.64]$).
During the study,
the participants were all curious about how to move the glyphs in the \emph{height} dimension
when we asked them to do so.
After showing our solutions,
most of the participants were impressed with
our approach for laying out objects in 3D space
because \emph{``it is much better than dragging the three axes on (the) 2D screen to move objects in 3D space.''} (P11).
They found these layout methods not only to be intuitive and efficient
but also \emph{``engaging''} (P3) and \emph{``fun to use''} (P4).

\begin{figure}[h]
	\centering 
	\vspace{-3mm}
	\includegraphics[width=1\columnwidth]{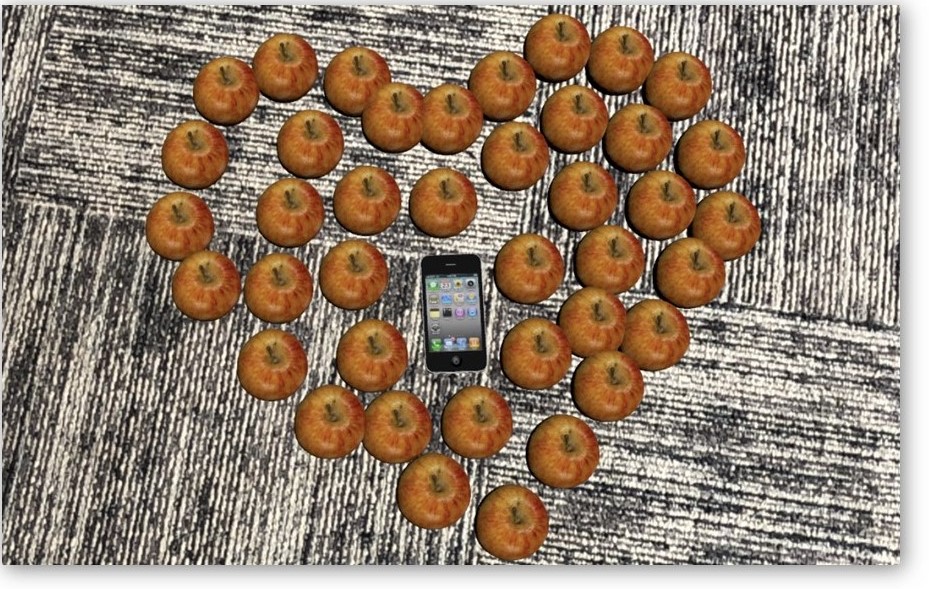}
	\vspace{-7mm}
	\caption{``Apples and Apple" designed by a participant in the third task, 
		depicting the money to buy an Apple phone can buy dozens of apples.
		One apple in the graphics represents a dozen apples.}
	\label{fig:user-generated}
	\vspace{-2mm}
\end{figure}

\textbf{Expressiveness.}
	Compared with the usability,
	there is still room for improvement when it comes to
	the expressiveness 
	with regards to the participants' rating in Q5,
	``I can create the ARGVis that I want'' ($\mu=5.82$, 95\% $CI = [5.37, 6.27]$).
	Seven participants pointed out that
	the inability of
	text input
	makes 
	MARVisT fail to allow adding graphics guides (\emph{e.g.}, title, axis, legend, etc.).
	We proposed to add voice input methods in the future 
	to support text input 
	while still avoiding keyboard input.
	All the participants were satisfied with our proposal.
	Another criticism about the expressiveness is 
	the lack of ability to create virtual glyphs,
	which is beyond the scope of this study and has been left for the future.
	Nevertheless, we observed that none of the participants created the same designs 
	though they were unembellished given the constraints of time, data, and the real-world environment (\emph{e.g.,} \autoref{fig:user-generated}).

\textbf{Usefulness.}
When considering the AR features that leverage information from reality
to facilitate the authoring,
the participants commented positively and confirmed their usefulness (Q6 - Q8).

The participants agreed that 
the context-aware nudging mechanism,
which can improve the usability,
was useful (Q6, $\mu=6.57$, 95\% $CI = [6.12, 7.02]$).
Most of the participants gave positive feedback,
and commented that 
\emph{``it provides stepping stones to help users initialize the design.''} (P13)
and \emph{``it is like a grammar checker, helping me to avoid errors.''} (P5)
We observed that only three participants
pointed out the perspective distortion problem
when encoding data with the size channels 
that have the most components on the depth axis.
This indicated our consideration of
providing design critiques based on both real-world background and design principles
is reasonable and necessary for non-experts.

As indicated by Q7, the participants 
spoke highly of how we utilize the information from reality to facilitate authoring visual scales,
and rated that the visual scales synchronization was useful ($\mu=6.86$, 95\% $CI = [6.67, 7.05]$).
For this feature,
the participants favored it because it not only \emph{``can dramatically improve the efficiency''} (P6)
but also \emph{``is impressive as I have never seen this (feature) before.''} (P9)
The intuitiveness of this feature 
is another reason the participants appreciated it,
\emph{e.g.}, \emph{``this feature is self-explanatory and the results
	(of interaction) are as expected.''} (P9)

In contrast, 
although the participants also favored the auto-layout method,
and rated that it was useful in Q8 ($\mu=6.71$, 95\% $CI = [6.46, 6.96]$),
they found it was not as intuitive as the visual scales synchronization feature.
P11 noted,
\emph{``this feature is useful especially when there are multiple objects to place,
	but it took me some time to understand how it works.''}
Comments indicated this was caused by the participants' lack of familiarity 
with the concept of creating mappings between objects through selecting attributes (\emph{JOIN}),
\emph{e.g.}, \emph{``...I am not a CS guy. I have never used this before.''} (P10)
However, we observed that once the participants got the idea of how it works,
they could quickly use it for rapidly placing glyphs with their physical referents.
Several participants expressed that the cost of time 
to learn this feature was acceptable 
for the benefit it brought, 
\emph{e.g.}, \emph{``Every time you want to use a powerful tool you need to learn something.''} (P14)

\textbf{Engagement.}
The participants also responded positively on visualizing information in AR form.
The rating on the statement of Q9 that 
\emph{``I enjoy the creation.''} ($\mu=6.71$, 95\% $CI = [6.39, 7.03]$)
suggested that MARVisT aroused their creative desires.
All the participants agreed that MARVisT worked in the right direction and 
AR techniques had great potential in presenting information,
especially in the personal context:
\emph{``I would like to use traditional charts when reporting data in a work context,
	but AR infographics is a better choice for personal context.''} (P13)

\textbf{Suggestions for Improvements.}
Additionally, 
the participants also provided constructive comments that imply further improvements.
A more powerful and intelligent template for initialization was wanted
because \emph{``sometimes it is hard to decide where to start.''} (P5)
For the limitation of creating virtual glyphs,
P8 suggested that MARVisT can allow users to 
create approximate glyphs
using the bounding box and texture of real objects
as \emph{``sometimes users do not need high fidelity 3D reconstruction models.''}
We also observed that the participants like to place virtual glyphs
in different places to see how they interact with real objects,
inspiring us to transfer functions like SnapToReality~\cite{SnapToReality}
to further improve the efficiency.
Another shortcoming that affected users' experiences is the performance of AR techniques, 
including the accuracy of tracking and registration.
Other minor issues such as missing Un-Do functions are ignored
as we do not aim to develop a full-featured system.

\vspace{-2mm}
\section{Discussion}
\label{sec:discussion}

\subsection{Mobile AR for Personal Visualization}
AR techniques are attracting considerable attention from the field of visualization
because of the popularity of low-cost AR devices,
especially mobile AR devices, 
which are one of the most accessible devices in our daily life.
In the wake of this surge of AR techniques,
we aim to explore the possibility of utilizing these methods in data visualization.
However, this potential technique is the topic of a heated debate.
On the one hand, AR visualization offers two inherent benefits, namely,
the engaged representation, which stands out from existing media,
and its context-preserving capability,
which seamlessly blends the digital information with the real-world context into a cohesive view.
On the other hand, visualizations in AR environments are criticized 
for the occlusion and distortion issues caused by their 3D form and rendering space,
which lead to inaccuracies in presenting data and errors in analytical tasks.

We make the observation that AR techniques are well suited for personal contexts,
in which accuracy is not the first priority~\cite{PV&PVA}, 
and other aspects, such as aesthetics and engagement, play important roles~\cite{p-engage}.
Utilizing AR techniques in the personal context
can fully leverage the advantages of AR for creative and engaging presentations
and circumvent the disadvantages of 3D visualizations that are not suitable for analytic purposes.
Moreover, as accessible all-in-one platforms,
mobile devices bridge the gap between AR techniques and general users,
bringing opportunities to make use of AR for PV practical.

Based on this observation,
we took the first step toward the use of mobile AR techniques in the personal context
by designing and developing MARVisT,
which enables non-experts to 
use mobile devices to
present their data together with the context 
that inspires their creations.
Although the infographics created by MARVisT 
are visually embellished by real-object glyphs,
which can have a negative impact on certain analytic tasks,
it is now generally understood that embellishment is not equivalent
to chart junk~\cite{DDG, usefuljunk}. 
Visual embellishments can help 
present and remember the context of data~\cite{usefulejunk2}
that makes it suitable for PV~\cite{isotype}.
The feedback from the user study
reflected participants' preferences of ARGVis for presentations and communication in the personal context.

We envision that AR techniques will bring considerable potential to PV
and mobile devices are more accessible than HMD devices for general users in the near future.
Thus, we hope that our work blazes a trail and arouses interest in applying mobile AR techniques for PV.
Although MARVisT is an authoring tool targeted at non-experts,
the three identified relationships between reality and virtuality
and the four proposed design considerations
can be generalized, 
and guide the design of advanced tools for other users.
Moreover, the interaction we designed
can be easily adapted to other mobile AR applications with various purposes.

\subsection{AR Visualization Authoring Tool}
Designing an AR visualization authoring tool
relies on research from several fields.
Essentially, it highly depends on the research on visualizing information in AR.
Visualizing information in AR 
and authoring AR visualization
have different challenges:
the former focuses on the challenges~\cite{AR-Canvas} like
\emph{How to avoid visual clutter and occlusion?};
\emph{What visualization types to choose?};
\emph{How to maintain faithful perception?} and the like;
while the latter considers
\emph{expressiveness},
\emph{efficiency},
and \emph{usability}, and so on.
Nevertheless,
research on visualization in AR can guide and foster
the design of AR visualization authoring tools.
Without further investigation into 
AR visualization,
many design decisions of MARVisT remain unclear.
For example,
when encoding data using the color channel,
which color we should use 
-- the material color or the screen color --
is still ambiguous.
Generally,
in traditional visualization,
we always use the screen color to encode data.
However, in AR environments,
the screen color of a virtual object is determined by several factors,
such as the object's material, and the lighting of the environment.
For high fidelity,
we should encode data using material color,
but it may result in a contrast effect~\cite{infovis-perception} that may bias users' perception.
Will this issue affect users' perception?
How can we address it?
Besides, 
the answers to questions about
creation pipeline and 
visualization types in AR
determine the \emph{expressiveness} of the authoring tool.
These open questions require further study.

In addition,
the design of authoring tools is influenced by 
progress in other research fields.
The \emph{Computer Vision} techniques 
can improve the \emph{efficiency} of the tool
and the connection between virtuality and reality.
For example, 
advanced \emph{Object Detection} techniques
can enhance the visual scales synchronization and the auto-layout features.
With the advancement of \emph{3D Scanning} techniques 
which allow the extraction of high quality 
and structural 
information
from the reality,
MARVisT could support 
users to reconstruct 3D glyphs from real objects
rather than fetch them from remote servers.
On the other hand,
the \emph{usability} of the tool mainly 
depends on
the user interfaces and interactions
on mobile devices,
which is an important topic in the HCI field.
MARVisT adapts UI designs of DataInk~\cite{DOD}, 
an authoring tool which supports data-driven sketching with pen and touch,
to support fluent interactions on mobile devices.
We believe a good authoring tool can encourage more attempts at AR visualization, 
which in turn fosters collecting requirements and designing better AR authoring tools.
With this work, we hope to encourage researchers to
pay attention to AR visualization,
eventually revealing the potential of this novel field.

\subsection{Limitations and Future Work}
The user study demonstrates that the participants
were able to create ARGVis with MARVisT in a short time
and were satisfied with MARVisT subjectively.
However,
these results should be treated carefully as well as qualitative studies with small sample size.
The participants also cannot fully represent the general public.
Further evaluations are warranted to confirm the results can be applied more generally.
Also, the results do not prove the interactions and UI designs of MARVisT are optimal
but only indicate that the overall designs have high usability.
We did not compare MARVisT with other tools
since currently, to the best of our knowledge,
we cannot find a similar tool as the benchmark.
Moreover,
the laboratory environment is not entirely equal to the personal context.
A field study is required to evaluate MARVisT in a real personal context.

Most of the limitations proposed by the users during the study were related to tool maturity.
For example, 
a text input function is wanted;
and 
functions which allow creating customized 3D models are missing.
MARVisT also involves certain inherent limitations,
which we consider as opportunities for future research.
First, users have to manually place the virtual glyphs
if they want to encode data attributes using position channels, 
or to place virtual glyphs to their physical referents
without creating the one-to-one mappings between them.
MARVisT should better support data binding on position channels to prevent incorrect positioning.
Second, MARVisT currently does not 
support
certain types of visualizations, 
such as node-link graphs and 
charts with axes.
Designing methods
that support these visualizations in mobile devices with limited screen size
need further work.
Third, although the primary disadvantages 
caused by the 3D 
are not as critical in PV as in visualizations serving analytics purposes,
these problems still trouble users when perceiving designs.
Fourth, the context-aware nudging mechanism is built upon heuristic rules 
and is not flexible enough.
We left the more powerful mechanism,
which may not only be built upon heuristic rules,
for future research.

\vspace{-2.5mm}
\section{Conclusion}
\label{sec:conclusion}
In this work, we present MARVisT, 
a mobile AR glyph-based visualization authoring tool tailored for non-experts.
MARVisT aims at utilizing the mobile AR techniques to 
tackle several unique challenges of PV
by enabling non-experts to use mobile devices 
for creating glyph-based visualizations in AR environments.
Our design of MARVisT is informed by 
existing principles and pipelines for authoring visualizations,
and a characterization of ARGVis
that help to identify the specificity of ARGVis authoring tools.
Through MARVisT, 
users can bind the visual channels to data attributes
and lay out virtual glyphs in 3D space
through direct manipulation and intuitive interactions.
MARVisT also 
leverages information from reality to assist non-experts in
visual mapping, visual scales synchronization, and auto-layout.
We demonstrate the expressiveness, usability, and usefulness of MARVisT through
a gallery of examples and a user study.

  \section*{Acknowledgments}
  The authors would like to thank Maps.com,
  the owner of the infographics Napoleon's Russian Campaigns,
  as well as the anonymous reviewers for their valuable comments.
  The work was supported by National Key R\&D Program of China (2018YFB1004300 ), NSFC-Zhejiang Joint Fund for the Integration of Industrialization and Informatization (U1609217), NSFC (61761136020, 61502416), Zhejiang Provincial Natural Science Foundation (LR18F020001) and the 100 Talents Program of Zhejiang University. 
  This project was also partially funded by RGC (CRF 16241916), Microsoft Research Asia, and the research fund of the Ministry of Education of China (188170-170160502).

\bibliographystyle{abbrv}
\bibliography{template}
%
%
\vskip -3\baselineskip plus -1fil
  \begin{IEEEbiography}[{\includegraphics[width=1in,height=1.2in,clip,keepaspectratio]{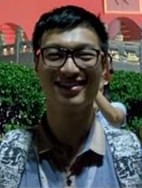}}]{Chen Zhu-Tian}
 	is currently working toward the Ph.D. degree in the Department of Computer Science and Engineering, Hong Kong University of Science and Technology. His research interests include AR visualization, immersive analytics, and visual analytics.
 	He received the bachelor's degree from the Department of Software Engineering, South China University of Technology in 2015.
 	For more information, please visit http://chenzhutian.org.
 \end{IEEEbiography}

 \vskip -3.5\baselineskip plus -1fil
  \begin{IEEEbiography}[{\includegraphics[width=1in,height=1.2in,clip,keepaspectratio]{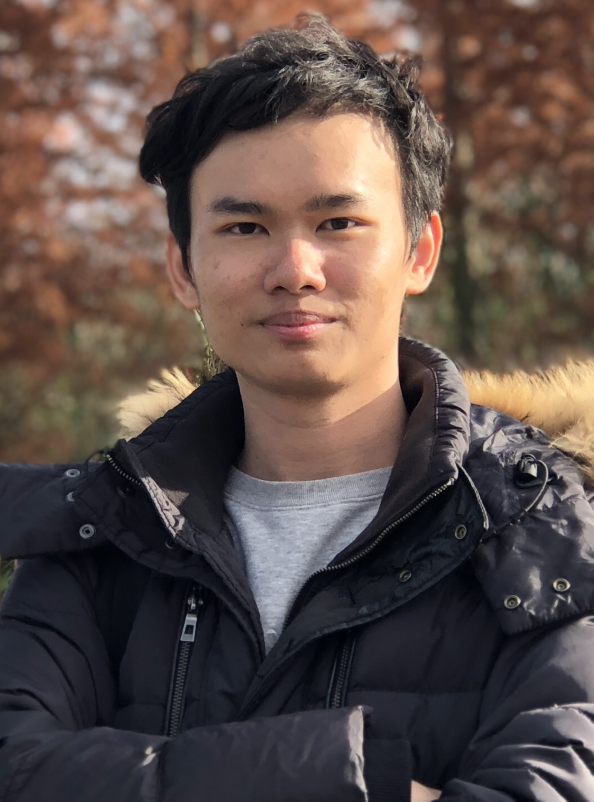}}]{Yijia Su}
 is currently a Research Assistant in the Department of Computer Science and Engineering at the Hong Kong University of Science and Technology (HKUST). He received his B.Eng degree in Software Engineering from South China University of Technology in 2016. His research interests include data visualisation and augmented reality techniques. For more information, please visit http://www.yijiasu.me/
  \end{IEEEbiography}

\vskip -3\baselineskip plus -1fil
  \begin{IEEEbiography}[{\includegraphics[width=1in,height=1.2in,clip,keepaspectratio]{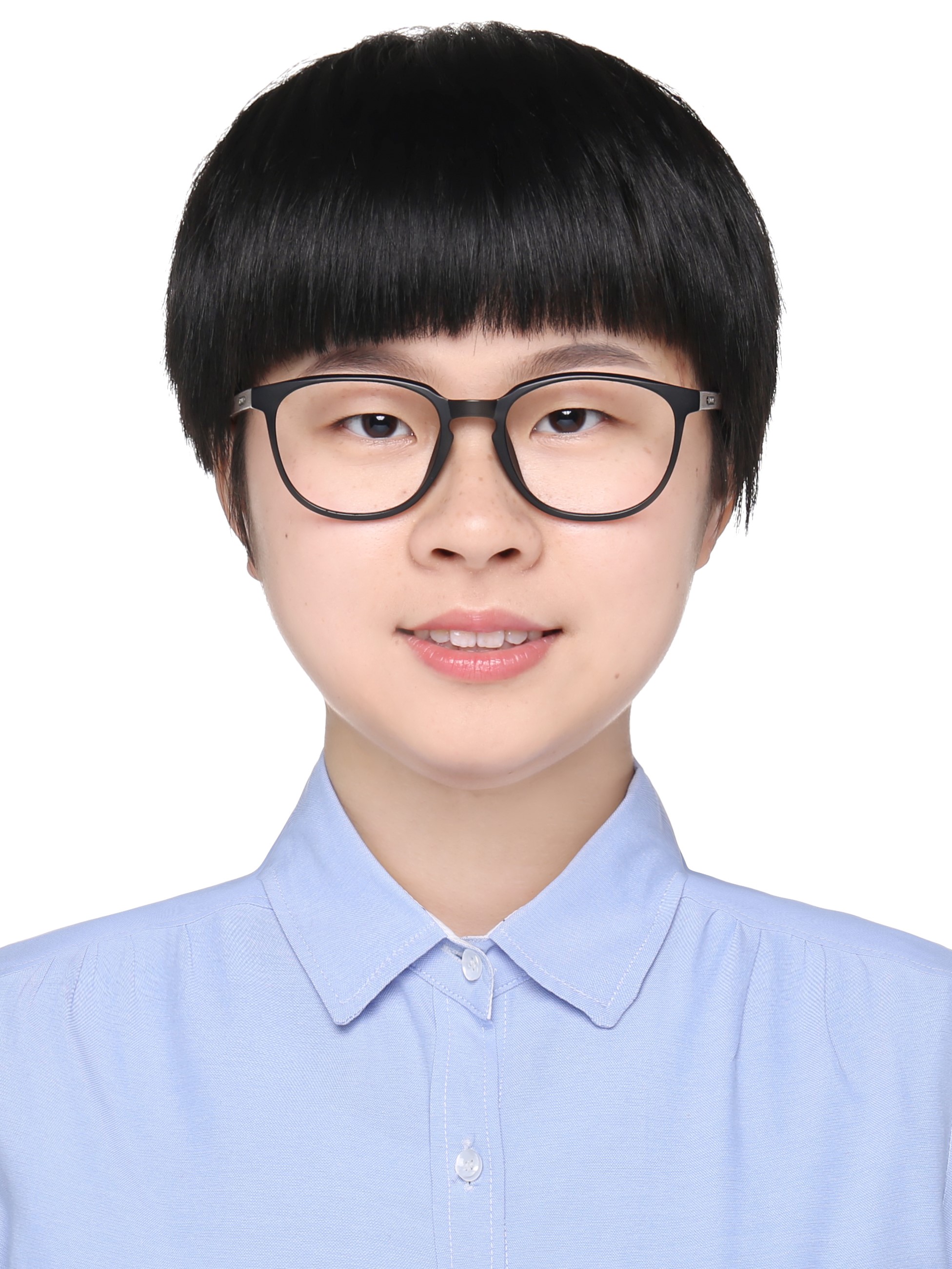}}]{Yifang Wang}
 	received her BEng degree in software engineering from Zhejiang University, China, in 2018. She will pursue her Ph.D. degree in the Department of Computer Science and Engineering at the Hong Kong University of Science and Technology (HKUST) from Sept. 2018. Her research interests include information visualization, visual analytics, and human computer interaction.
 \end{IEEEbiography}
 
 \vskip -3\baselineskip plus -1fil
  \begin{IEEEbiography}[{\includegraphics[width=1in,height=1.2in,clip,keepaspectratio]{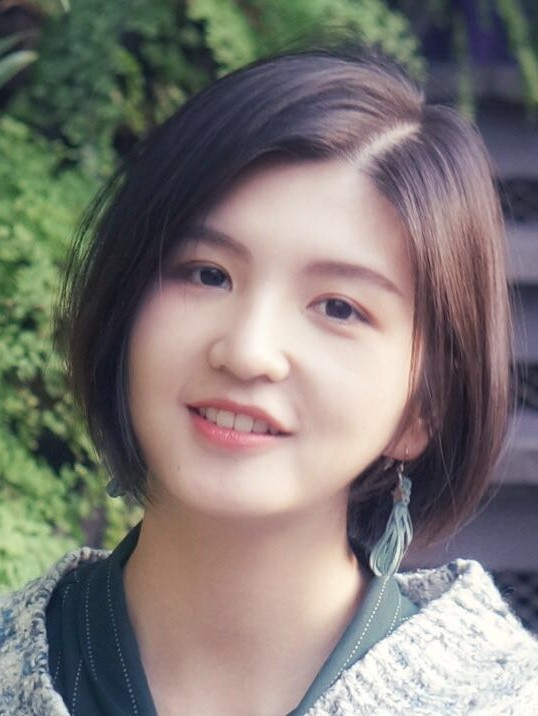}}]{Qianwen Wang}
 	is now a Ph.D. student at Hong Kong University of Science and Technology.
 	Her research interest focus on information visualization and interactive machine learning.
 	She received a BS degree in Electronics Engineering, Xi'an Jiaotong University.
 \end{IEEEbiography}

  \vskip -2.5\baselineskip plus -1fil
  \begin{IEEEbiography}[{\includegraphics[width=1in,height=1.2in,clip,keepaspectratio]{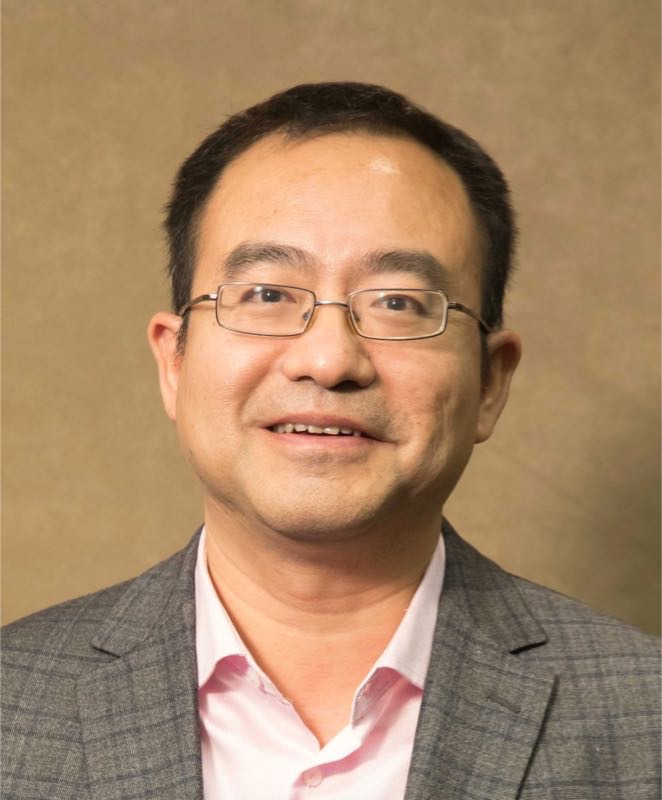}}]{Huamin Qu}
 	is a professor in the Department of
 	Computer Science and Engineering at the Hong
 	Kong University of Science and Technology. His
 	main research interests are  in visualization and human-computer interaction, with focuses on urban informatics, social network analysis, e-learning, text visualization, and explainable artificial intelligence. He obtained a BS in Mathematics
 	from Xi'an Jiaotong University, China, an
 	MS and a PhD in Computer Science from the
 	Stony Brook University. For more information,
 	please visit http://www.huamin.org.
 \end{IEEEbiography}

 \vskip -3\baselineskip plus -1fil
 \begin{IEEEbiography}[{\includegraphics[width=1in,height=1.2in,clip,keepaspectratio]{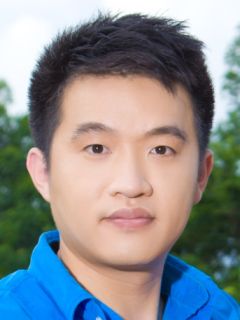}}]{Yingcai Wu}
   is a ZJU100 Young Professor at the State Key Lab of CAD\&CG, Zhejiang University. His main research interests are in information visualization and visual analytics, with focuses on user urban informatics, social media analysis, and sports science. He received his Ph.D. degree in Computer Science from the Hong Kong University of Science and Technology. Prior to his current position, Dr. Wu was a postdoctoral researcher in the University of California, Davis from 2010 to 2012, and a researcher in Microsoft Research Asia from 2012 to 2015. For more information, please visit http://www.ycwu.org.
\end{IEEEbiography}







\end{spacing}
\end{document}


\begin{spacing}{0.98}
%
\title{MARVisT: Authoring Glyph-based Visualization in Mobile Augmented Reality}
%
%
%
%

\author{Chen Zhu-Tian, 
	    Yijia Su,
	    Yifang Wang,
	    Qianwen Wang,
	    Yingcai Wu,
	    Huamin Qu
	    
 \IEEEcompsocitemizethanks{
	\IEEEcompsocthanksitem Chen Zhu-Tian, Yijia Su, Yifang Wang, Qianwen Wang and Huamin Qu are with Hong Kong University of Science and Technology, Hong Kong. 
    	E-mail: \{zhutian.chen, yijiasu, yifangwang, qwangbb\}@connect.ust.hk, huamin@cse.ust.hk.
   	\IEEEcompsocthanksitem Yingcai Wu is with State Key Lab of CAD\&CG, Zhejiang University, Hangzhou, China. E-mail: ycwu@zju.edu.cn.
 }
	    
}

%
%

\markboth{Journal of \LaTeX\ Class Files,~Vol.~14, No.~8, August~2015}%
{Shell \MakeLowercase{\textit{et al.}}: Bare Demo of IEEEtran.cls for Computer Society Journals}
\maketitle

\IEEEdisplaynontitleabstractindextext

%
\IEEEpeerreviewmaketitle

\appendices
\section{Explanation of technical details}
We present an example to demonstrate the technical details of
\textbf{Visual Scales Synchronization} and \textbf{Virtual Glyphs Auto-layout}
and how a designer 
can use MARVisT to create a keyboard frequency sculpture
in a few minutes.

\subsection{Visual Scales Synchronization}
\label{sec:vss}

\begin{figure}[h]
	\centering 
	\includegraphics[width=1\columnwidth]{ap-keyboard-case-sync}
	\vspace{-6mm}
	\caption{Four steps to synchronizing the visual scales between real objects and virtual glyphs. 
		The blue steps (Step 1, 2, and 4) are finished by MARVisT. 
		The yellow step (Step 3) is done by the user.}
	\label{fig:kb-sync}
\end{figure}

After encoding the typing frequency using the height and color of a bar (in \autoref{fig:kb-sync}),
the designer needs to adjust the size of a bar to be the same as that of a keycap.
Based on the basic interactions introduced in Section 4.2 of the paper,
the designer has to manually modify the size of a bar in a trial and error manner.
MARVisT provides an advanced method to help users finish this kind of task 
(i.e., synchronize visual scales between virtual glyphs and real objects) in four steps:

\begin{enumerate}[leftmargin=*]
	\vspace{-2mm}
	\item \textbf{Detect real objects}. MARVisT leverages several object detection methods 
	provided by ARKit to recognize real objects in the current camera images as much
	as possible. Specifically, the methods we used include 3D object detection~\cite{AppleA}, image detection~\cite{AppleB}, and object detection based on Vision frame work~\cite{AppleC, AppleD} (wherein the model was trained in Turi Create using Darknet YOLO). 
	All of these methods are embedded in ARKit so it is easy to use them by calling their APIs. 
	Once a real object is successfully detected, 
	it will be highlighted with a flicker effect (in \autoref{fig:kb-sync}a). 
	All the detected real objects will be stored and passed to the next step. 
	We further elaborate the details of this step in \autoref{fig:code1} 
	in the form of a JavaScript-style pseudo code with documentation.
	\begin{figure}[h]
		\centering 
		\includegraphics[width=1\columnwidth]{ap-code1-light}
		\vspace{-6mm}
		\caption{The JavaScript-style pseudo code of the detection of real objects.
			The pseudo code elaborates the main process of how this function works.}
		\label{fig:code1}
		\vspace{-2mm}
	\end{figure}
	
	\item \textbf{Extract visual channels.} After recognizing real objects in the current camera images, MARVisT will try to extract the visual channels of each real object as much as possible. Specifically, the \emph{position} channels (\emph{x, y, z}) in the world coordinate of the AR environment can be extracted once the real object is detected; the \emph{size} channels (\emph{1D-length, 2D-area, 3D-volume}) are estimated based on the real object's bounding box, which is detected in the previous step; the \emph{text} channel is detected and extracted using the Vision framework~\cite{AppleE} provided by Apple and only the text with the largest area will be used. Not all the visual channels can always be extracted. MARVisT will display the available visual channels when the user taps on a detected real object (in \autoref{fig:kb-sync}b). We further elaborate the details of this step in \autoref{fig:code2} in the form of the JavaScript-style pseudo code with documentation.
	\begin{figure}[h]
		\vspace{-4mm}
		\centering 
		\includegraphics[width=1\columnwidth]{ap-code2-light}
		\vspace{-8mm}
		\caption{The pseudo code elaborates the main process of how to extract the visual channels of real objects.}
		\label{fig:code2}
		\vspace{-2mm}
	\end{figure}
	
	\item \textbf{Assign visual channels.} This step is finished by the user. When the user taps on a detected real object, a single-ring semi-annulus similar to the one of the virtual glyphs will pop up. The semi-annulus (in Figure.1a) consists of beads which represent the visual channels of the real objects. The user can drag a bead and drop it on a virtual glyph to assign its value to the corresponding visual channel of the virtual glyph (in \autoref{fig:kb-sync}c).
	
	\item \textbf{Synchronize visual scales.} If the visual channel has not been used to encode data attributes, MARVisT will automatically assign the value of the visual channel of the real object to all virtual glyphs of the same type. If this visual channel has already been used to encode a data attribute, MARVisT will inversely calculate the new scale based on the value of the real object's visual channel and the data bounded with the virtual glyph. Then the new scale will automatically be propagated to all virtual glyphs of the same visual mapping, leading to updates of the corresponding virtual glyphs' visual channels (in \autoref{fig:kb-sync}d). We further elaborate the details of this step in \autoref{fig:code3} in the form of a JavaScript-style pseudo code with documentation.
	\begin{figure}[h]
		\vspace{-3mm}
		\centering 
		\includegraphics[width=1\columnwidth]{ap-code2-light}
		\vspace{-8mm}
		\caption{The pseudo code elaborates the main process of how the final step synchronizes visual scales.}
		\label{fig:code3}
		\vspace{-6mm}
	\end{figure}

\end{enumerate}

\subsection{Virtual Glyphs Auto-layout}
\begin{figure}[h]
	\vspace{-2mm}
	\centering 
	\includegraphics[width=1\columnwidth]{ap-keyboard-case-autoLayout}
	\vspace{-6mm}
	\caption{Steps for auto-layout. The first and second steps are the same as those in \autoref{fig:kb-sync}.
		a) Tap on any keycap to open the semi-annulus of a real object and tap on any 3D bar to open the semi-annulus of a virtual glyph. 
		b) Map the keycaps to 3D bars based on the text channel of the keycaps and the name attribute of the 3D bars. 
		c) Automatically place each 3D bar onto its corresponding keycap. }
	\label{fig:kb-autolayout}
\end{figure}

After adjusting the size of the bars to the size of keycaps,
the designer needs to place each bar onto its corresponding keycap.
It is quite tedious for the user to manually move each virtual bar onto
its physical referent. MARVisT provides an automated method to
help users finish this kind of task in four steps:
\begin{enumerate}[leftmargin=*]
	\item \textbf{Recognize real objects}. This step is the same as the one introduced in \autoref{sec:vss} and can reuse the results.
	
	\item \textbf{Extract visual channels.} This step is the also same as the one introduced in \autoref{sec:vss} and can also reuse the results.
	
	\item \textbf{Map visual channels}. This step is done by the user. To map real objects to virtual glyphs, MARVisT supports the user to use visual channels as the mapping key. For example, the user wants to map the 3D bars, which encode the typing frequency with the height and color channels, to their corresponding keycaps. After detecting the keycaps from the keyboard, MARVisT can extract several visual channels of the keycaps, such as width, height, and text. The user can simultaneously open the semi-annulus of the virtual glyphs and the real objects (in \autoref{fig:kb-autolayout}a). Then the user can drag the bead of the selected visual channel of the real objects (e.g., the text channel) and drop it onto the bead of the selected data attribute (e.g., the name of the keycap) of the virtual glyphs to specify the mapping relationship (in \autoref{fig:kb-autolayout}b). In \autoref{fig:kb-autolayout}, a keycap will be mapped to the 3D bar whose name is equal the text channel on the keycap.
	
	\item \textbf{Lay out virtual glyphs}. After the user specifies the mapping between real objects and virtual glyphs, MARVisT will automatically place each virtual glyph to its corresponding real object (in \autoref{fig:kb-autolayout}c), whose position in the world coordinate of the AR environment is known after being detected. The details of this step are elaborated in \autoref{fig:code4} in the form of a JavaScript-style pseudo code  with documentation.
	\begin{figure}[h]
		\vspace{-4mm}
		\centering 
		\includegraphics[width=1\columnwidth]{ap-code4-light}
		\vspace{-8mm}
		\caption{The pseudo code elaborates the main process of how MARVisT places virtual glyphs to their physical referents automatically after the user has specified the mappings between them.
		}
		\label{fig:code4}
	\end{figure}
\end{enumerate}

\subsection{Supplemental Details}
For the examples in the paper,
we provide more details of
the ping pong ball example of the visual scales synchronization in \autoref{fig:pingpong}
and the sugar stack example of the virtual glyphs auto-layout in \autoref{fig:drinks}.

\begin{figure}[h]
	\centering 
	\includegraphics[width=1\columnwidth]{ap-pingpong}
	\vspace{-6mm}
	\caption{
		a) After tapping on the ping pong ball, a single-ring semi-annulus which consists of beads with a blue border will pop up. 
		b) The user can drag the bead out of the semi-annulus and drop it onto a virtual glyph to assign its value to the corresponding visual channel of the virtual glyph. 
		c) MARVisT automatically synchronizes the corresponding visual channel of the virtual glyph. 
	}
	\label{fig:pingpong}
\end{figure}

\begin{figure}[h!]
	\centering 
	\includegraphics[width=1\columnwidth]{ap-drinks}
	\vspace{-6mm}
	\caption{
		a) Detect the three drinks and extract their size channels based on their bounding boxes. 
		b) Open the semi-annulus panel to reveal the visual channels of a real object and the data attributes a virtual glyphs. 
		c) Map the drinks to the sugar stacks based on the volume channel and the sugar content attribute: the greatest volume to the most sugar content, and so on. 
		d) MARVisT automatically places each sugar stack in front of its corresponding drink.}
	\label{fig:drinks}
	\vspace{-2mm}
\end{figure}

\section{Performance Testing}
\label{sec:performtesting}
We conducted experiments to evaluate the performance
of the current implementation of MARVisT. We evaluated 
the construction times, 
the frame rates of the static scenes, 
and the frame rates of the dynamic scenes for varying data sizes and glyph
model complexities on an iPhone 8 plus (CPU with 4 processors
@ 2.34GHz + 2 processors @ 1.7GHz, Apple A11 GPU, 3GB RAM)
and an iPad Pro (CPU with 4 Vortex + 4 Tempest , Apple A12x GPU, 4GP RAM).
The number of glyphs ranged from 10 to 1000 to 10000.
We used two built-in primitive models, namely, cube and sphere,
and the house and shoes models, which is the same model we used in Figure.5 in the paper,
to represent simple and complex glyph models respectively (\autoref{fig:models}).
Each time we generated the glyphs and randomly distributed them within the field of view.

\begin{figure}[h]
	\centering 
	\includegraphics[width=1\columnwidth]{ap-models}
	\vspace{-6mm}
	\caption{Simple: Cube and Sphere. Complex: Shoe and House.}
	\label{fig:models}
	\vspace{-2mm}
\end{figure}

\begin{figure}[h]
\centering 
\includegraphics[width=1\columnwidth]{ap-test1}
\vspace{-6mm}
\caption{The construction time of simple (cube and sphere) and complex (shoe and house)
models running on an iPhone 8 Plus and an iPad Pro. The y-axis indicates the time for construction,
and the x-axis indicates the data size. Given the memory limitations of the devices, we are not
able to load 10, 000 complex models on an iPhone 8 Plus and an iPad Pro.}
\label{fig:test1}
\vspace{-2mm}
\end{figure}

To measure the \textbf{construction time}, we imported the data and
rendered the glyphs 11 times for each test case. 
We skipped the first time as a warm-up and report the
average of the remaining 10 (in \autoref{fig:test1}). Even with 10, 000
models construction times remained below 12 seconds on
both the devices. Given the memory limitations of the two
devices (3GB on iPhone 8 Plus and 4GB on iPad Pro), we could
not load and render 10, 000 complex models (36kb for each shoe
and 60kb for each house).

\begin{figure}[h]
\centering 
\includegraphics[width=1\columnwidth]{ap-test2}
\vspace{-6mm}
\caption{The frame rates of static simple (cube and sphere) and complex (shoe and house)
models running on an iPhone 8 Plus and an iPad Pro. The y-axis indicates the frame rates,
and the x-axis indicates the data size. Given the memory limitations of the devices,
we are not able to load 10, 000 complex models on an iPhone 8 Plus and an iPad Pro.}
\label{fig:test2}
\vspace{-2mm}
\end{figure}

To measure the \textbf{frame rates of the static scenes},
we displayed all glyphs within the field of view for one minute. 
The maximum frame rates on the two
devices were 60 FPS given the hardware restrictions.
As expected, the frame rates dropped along with
the increasing complexity of the scene (in \autoref{fig:test2}). 
The frame
rate of the house model on iPhone 8 Plus dropped quickly because
the mechanism of iOS\footnote{https://developer.apple.com/documentation/scenekit/scnview/1621205-preferredframesperscond} 
restricted the maximum frame rate to 30 FPS
under heavy workloads.

\begin{figure}[h]
\centering 
\includegraphics[width=1\columnwidth]{ap-test3}
\vspace{-6mm}
\caption{The frame rates of dynamic simple (cube and sphere) and complex (shoe and house)
models running on an iPhone 8 Plus and an iPad Pro. The y-axis indicates the frame rates,
and the x-axis indicates the data size. Given the memory limitations of the devices,
we are not able to load 10, 000 complex models on an iPhone 8 Plus and an iPad Pro.}
\label{fig:test3}
\vspace{-3mm}
\end{figure}

To measure the \textbf{frame rates of the dynamic scenes}, we displayed all glyphs
within the field of view and used the data-binding panel to increase
their volume to twice the size. The maximum frame rates on the two
devices were 60 FPS given the hardware restrictions.
Compared with the frame rates of static scenes,
the frame rates of the scene wherein the models are dynamically changed
dropped a little bit as expected. Besides that, the frame rates dropped along with
the increasing number of models in the scene (in \autoref{fig:test3}).

Overall, for datasets with reasonable size (1, 000 models or less), 
our implementation 
guarantee fast construction (in around 2 seconds)
and real-time frame rates (over 50 FPS in most cases).
Given the hardware and software limitations
of the two iOS devices, we could not load 10, 000 data points and the
frame rates drop quickly in heavy workloads. However, considering the
usage scenario (i.e., the personal context) of MARVisT, wherein users
usually do not have big datasets to visualize, we think the performance
of the current implementation is acceptable. 
We believe there is still room for
improvement of MARVisT in the future, such as optimizing the memory usage,
utilizing level-of-detail techniques to improve the frame rate, and following
the best practices of iOS applications to enhance the overall performance of MARVisT.

\bibliographystyle{abbrv}
\bibliography{template}
\end{spacing}